\documentclass[structabstract,printer]{aa}
\usepackage{amsmath,longtable,lscape,natbib}   
\pdfoutput=1
\usepackage[pdftex]{graphicx}  
\usepackage{verbatim}
\usepackage{url}

\usepackage{amssymb}  

\begin{document}


\title{Cosmic-ray-induced ionization in molecular clouds adjacent to supernova remnants}
\subtitle{\textbf{Tracing the hadronic origin of GeV gamma radiation}}
\author{Florian Schuppan\inst{1}\fnmsep\thanks{Corresponding author. Contact: florian.schuppan@rub.de, phone: +49-234-3222329} \and Julia K.~Becker\inst{1} \and John H.~Black\inst{2} \and Sabrina Casanova\inst{1,3,4,5}}
\institute{Ruhr-Universit\"at Bochum, Fakult\"at f\"ur Physik \&
  Astronomie, 44780 Bochum, Germany \and Dept. of Earth and Space Sciences, Chalmers University of Technology, 
  Onsala Space Observatory, SE-439 92 Onsala, Sweden \and Unit for Space Physics, North-West University, Potchefstroom 2520, South Africa \and Universit\'{e} Paris Diderot-Paris 7, Laboratoire APC, 
  B\^{a}timent Condorcet, Case 7020, 75205 Paris Cedex 13 \and Max-Planck-Institut f\"ur Kernphysik, Saupfercheckweg 1, 
  69117 Heidelberg, Germany}    

\date{\today}

\abstract
   {Energetic gamma rays (GeV to TeV photon energy) have been
    detected toward several supernova remnants (SNR) that are associated
		with molecular clouds. If the gamma rays are
		produced mainly by hadronic processes rather than leptonic processes
		like bremsstrahlung, then the flux of energetic cosmic ray nuclei ($>1$~GeV)
		required to produce the gamma rays can be inferred at the site
		where the particles are accelerated in SNR shocks. It is of great
		interest to understand the acceleration of the cosmic rays
		of lower energy ($<1$~GeV) that accompany the energetic component.
		These particles of lower energy are most effective in ionizing
		interstellar gas, which leaves an observable imprint on the
		interstellar ion chemistry. A correlation of energetic gamma
		radiation with enhanced interstellar ionization can thus be used to
		support the hadronic origin of the gamma rays and to constrain the
		acceleration of ionizing cosmic rays in SNR.}
   {We propose a method to test the hadronic origin of GeV gamma rays
    from SNRs associated with a molecular cloud.}
   {We use observational gamma ray data for each SNR known to 
    be associated with a molecular cloud, modeling the observations to 
    obtain the underlying proton spectrum under the assumption that the gamma 
    rays are produced by pion decay. Assuming that the acceleration mechanism does 
    not only produce high energy protons, but also low energy protons, this proton 
    spectrum at the source is then used to calculate the ionization rate of 
    the molecular cloud. Ionized molecular hydrogen triggers a chemical network 
    forming molecular ions. The relaxation of these ions results
    in characteristic line emission, which can be predicted.}
   {We show that the predicted ionization rate for at least two objects is more than an
    order of magnitude above Galactic average for molecular clouds, hinting
    at an enhanced formation rate of molecular ions. There will be interesting 
    opportunities to measure crucial molecular ions in the infrared and 
    submillimeter-wave parts of the spectrum.}
   {}

\authorrunning{F.\ Schuppan et al.}
\titlerunning{CR induced ionization in SNR-adjacent molecular clouds}   
\maketitle

\section{Introduction\label{intro}}

The origin of cosmic rays (CRs) is an open question in astrophysics. The cosmic ray spectrum below the knee, at energy/nucleon $E<10^{15}$~eV is believed to be associated with cosmic ray acceleration in supernova remnants (SNRs) \citep{bell1978,blandford1978,blandford1987}. However, there is no conclusive proof for this until now. There is an excess in GeV-TeV gamma rays observed from SNRs associated with molecular clouds, see e.g.\ \cite{abdo(W51C)2009}, \cite{abdo(W44)2010} and \cite{aharonian2008}. These signals might be caused by bremsstrahlung or inverse Compton scattering of electrons in a leptonic scenario, or by the decay of neutral pions formed by proton-proton scattering in a hadronic scenario. So far, it is not known which of these processes is dominant. Investigating which is the dominant process is important to understand the origin of cosmic rays. 
In the hadronic scenario, high energy protons are accelerated in the SNR shocks and then escape to interact with ambient protons, in particular in molecular clouds in the direct vicinity of the SNR. It is also likely that low energy protons ($E<1$~GeV) are accelerated in the SNR, but these protons fall below the threshold for pion formation, so that no conclusions concerning the low energy CR spectrum can be drawn from gamma ray observations. However, low energy protons are very efficient in ionizing molecular gas. Therefore, ionization signatures provide information about the density of low energy cosmic rays. 
The main product of the ionization of molecular hydrogen, H$_2^+$, initiates a chain of chemical reactions that yield additional ions like H$_3^+$, OH$^+$, H$_2$O$^+$, H$_3$O$^+$ and HeH$^+$ \citep{black2007,mccarthy2006}.
These molecules are most likely formed in rotationally and vibrationally excited states, the corresponding wavelength for relaxation in the UV or IR, respectively. If the abundances of these molecules are sufficiently large, the UV or IR signals might be detectable and offer conclusions concerning the source of cosmic rays.
A correlation study of molecular clouds bright in GeV gamma rays and showing ionization features might be useful to find the dominant process in forming GeV gamma rays.
In this paper, the ionization rate of molecular hydrogen is calculated for each SNR known to interact with a molecular cloud. In contrast to former work \citep{becker2011}, here the spectral shape of the primary particle spectral energy distribution (SED) below kinetic energies of $E$~$\sim$~1~GeV is altered in order to take the unknown spectral behavior into consideration, rather than altering the minimum energy of particles contributing to the ionization process.
The paper is structured as follows: In section \ref{proc} the competing processes active while particles are being accelerated
and their influence on the spectral shape of the CRs is discussed, in section \ref{prim_SED} the spectrum of the primary protons is calculated by considering loss processes and the acceleration mechanism, in section \ref{ion_rate} the ionization rate for each SNR associated with a molecular cloud is calculated, in section \ref{uncer} the uncertainties are discussed, in section \ref{signatures} the ionization signatures to be expected are shown, in section \ref{observations} first observational evidence for an enhanced ionization rate in correlation with GeV gamma ray emission is summarized and in section \ref{conclusions}, a summary of the paper as well as an outlook to future work is given.


\section{Acceleration and diffusion\label{proc}}

Since the primary particle spectra at energies below $\sim$~1~GeV at
the source are not known, especially in the context of cosmic
ray-induced ionization (e.g.\ \cite{nath1994}, \cite{indriolo2009}),
the competing processes affecting these particles are discussed and
compared in this section. In particular, it is of importance at what
energy the ionization timescale is shorter than the acceleration timescale and vice versa to ensure that acceleration is unaffected. The acceleration timescale is given in \cite{jokipii1987}, \cite{biermann-acc} as
\begin{equation}
	\tau_{\rm acc} = \frac{8 \kappa}{V_{\rm sh}^2}~,
\end{equation}
where $\kappa$ is the diffusion coefficient and $V_{\rm sh}$ is the shock velocity. The diffusion coefficient may well differ inside the cloud and outside the cloud. Outside the cloud, the diffusion coefficient has to be low enough to allow for efficient acceleration, while inside the cloud the diffusion coefficient has to be sufficiently large for the particles to penetrate the cloud within the age of the SNR. For a typical age of $T=10^4$~yr and a penetration depth of $R=30$~pc \citep{becker2011}, the diffusion coefficient for a particle of $p~=~1$~GeV~c$^{-1}$ would have to be $\kappa~=~\frac{1}{2}\frac{R^2}{T}~\approx~1.4\times10^{28}$~cm$^2$~s$^{-1}$. Introducing a momentum dependence in the diffusion coefficient, $\kappa~=~\kappa_0~\left(\frac{p}{{\rm 1~GeV}~c^{-1}} \right)^{\delta}$, and applying this value for the cloud's interior, the acceleration timescale can be written as
\begin{equation}
  \tau_{\rm acc} = 4.4\times 10^{13} \left(\frac{{500\,{\rm km\ s}^{-1}}}
{{V_{\rm sh}}}\right)^2 \left(\frac{p}{{1\,{\rm GeV\ c}^{-1}}}\right)^{\delta}
\;\; {\rm s}~,
\end{equation}
where $c$ is the speed of light and $p$ is the momentum of the particle. This
includes a scaling to a typical value of the shock velocity for middle-aged remnants, $V_{\rm sh}=500$ km s$^{-1}$ \citep{abdo(W44)2010}.

\begin{figure}
	\centering
		\includegraphics[scale=0.55]{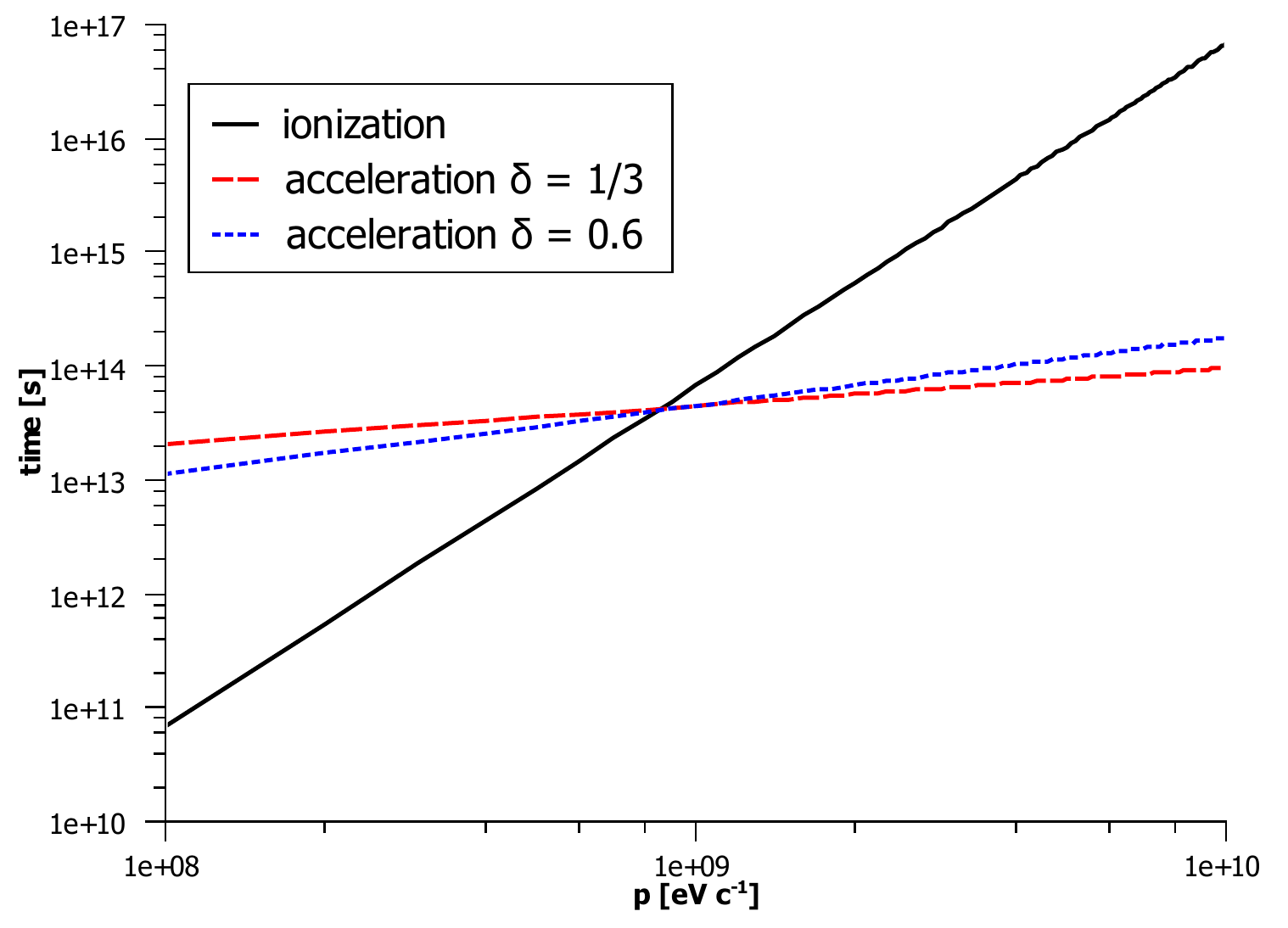}
	\caption{Ionization and acceleration timescale for protons, $n$~=~100~cm$^{-3}$ and $V_{\rm sh}$~=~500~km/s.}\label{fig:timescales}
\end{figure}

The momentum loss rate ${{\rm d}p}/{{\rm d}t}$ at a given momentum $p$ by ionization or excitation is given by \cite{lerche-schlick1982} as
\begin{equation}
\begin{aligned}
	\frac{{\rm d}p}{{\rm d}t} = -5 \times 10^{-19}~ q^2 \left(\frac{n}{\rm cm^{-3}} \right)\left(\frac{p}{m c} \right)^{-2} \\ \cdot \left[11.3 + 2\cdot \ln{\left(\frac{p}{m c} \right)} \right]~{\rm eV~cm^{-1}}~,
\end{aligned}	
\end{equation}
where $q$ is the charge of the particle, $n$ is the number density of the interacting region, $m$ is the mass of the particle and $c$ is the speed of light. A more general expression can be found in \cite{mannheim1994}. Neglecting the logarithmic dependence in the square bracket, since $p$ is of the order of $\sim1$~GeV~$c^{-1}$, the ionization timescale for protons is calculated as
\begin{equation}
	\tau_{\rm ion} = \left(\frac{{\rm d}p}{{\rm d}t} \right)^{-1} \cdot p = 5.6 \times 10^{15}~ ~q^{-2}\left(\frac{n}{\rm cm^{-3}} \right)^{-1} \left(\frac{p}{m c} \right)^3~{\rm s}~.
\end{equation}
These two timescales are compared in Fig.~\ref{fig:timescales}, where $q$~=~1, $n$~=~100~cm$^{-3}$ and $V_{\rm sh}$~=~500~km/s were used as a typical set of parameters. The momentum dependence of the diffusion coefficient is shown for $\delta~=~1/3$ as well as for $\delta~=~0.6$, as discussed in \cite{blasi2011} and references therein. There it is reported that a value of $\delta~=~1/3$ would favor second order Fermi acceleration and at the same time explain the observed ratios of B/C and fit the anisotropy of cosmic rays observed at Earth better than $\delta~=~0.6$. However, this would require an injection spectrum of $N(E)~\propto~E^{-2.4}$, which is challenging for non-linear diffusive shock acceleration (NLDSA) in SNRs. On the other hand, $\delta~=~0.6$ would favor NLDSA and match the detected spectra of cosmic rays including nuclei heavier than helium, but result in an anisotropy larger than observed. The shock velocity $V_{\rm sh}$~=~500~km/s is typical for older SNRs, as e.g.\ W51C \citep{abdo(W51C)2009}. For younger SNRs, the shock velocity can reach values of $V_{\rm sh}~\sim~$10,000~km/s.

In the environment described by the chosen parameters, at a particle momentum of $p~\geq$~0.8~GeV~$c^{-1}$, almost independent of the actual momentum dependence of the diffusion coefficient, the acceleration timescale is shorter than the ionization timescale, indicating that ionization losses do occur, but do not suppress the acceleration process effectively above this momentum. Therefore, the ionization losses do affect the spectral index of the primary proton SED significantly, but only at momenta $p$~$\leq$~1~GeV~$c^{-1}$.

Furthermore, adiabatic deceleration might in principle occur and alter the spectral shape of the primary proton SED, especially for momenta $p$~$\geq$~1~GeV~$c^{-1}$ (see \cite{lerche-schlick1982}). The momentum loss for primary particles with momenta $p$~$\geq$~1~GeV~$c^{-1}$ by adiabatic deceleration dominates the ionization losses, while below this energy the ionization losses dominate losses by adiabatic deceleration (see \cite{lerche-schlick1982}).
However, the primary proton SED at these energies is obtained from modeling the GeV gamma ray emission via $\pi^0$-decay using the Kamae model \citep{kamae2006,karlsson2008}. The lowest observable photon energy $E_{\gamma}~\approx~$100~MeV corresponds to a primary proton energy of $E_{\rm p}~=$~1~GeV. Thus, above 1~GeV the spectrum of primary protons needed is directly known and would already include modulation effects on the SED caused by adiabatic deceleration. Yet, there is no direct information about the low energy cosmic rays, so the estimate of the low energy cosmic ray spectrum has to be made very carefully.


\section{Primary proton SED \label{prim_SED}}

To calculate ionization by cosmic ray protons in a molecular cloud, the cosmic ray proton spectrum at the cloud is required. There is no observational data of the particle spectrum at the source: only the spectrum after propagation to Earth is known. Due to larger uncertainties in the description of the propagation process, especially due to the lack of knowledge of the exact magnetic field configuration, the spectrum at the source cannot be described easily from the observed data. If the magnetic field is known, it can be taken into consideration following \cite{padovani2011}. Gamma ray emission from hadronic interactions is, on the other hand, very well suited to derive the primary particle spectrum above $\sim$~1~GeV, since the gamma spectrum follows the primary spectrum. This gamma ray emission was detected e.g.\ by the \textit{Fermi}-LAT instrument. Assuming that the detected gamma radiation is mainly caused by the decay of neutral pions from inelastic proton-proton interactions at the cloud, the primary proton SED can be found modeling the gamma ray detections. The spectral shape of this SED is gained modeling the gamma ray emission from neutral pion decay, induced by inelastic proton-proton interactions, and fitting the modeled gamma ray spectrum to the observational data. Loss processes for the primary particles are summarized in subsection \ref{loss_proc}, acceleration processes are discussed in subsection \ref{acc_mech} and finally the calculation of the normalization of the primary proton SED is described in detail in subsection \ref{norm}.

\subsection{Loss processes\label{loss_proc}}

The primary protons accelerated by the SNR can suffer momentum losses on their way to the molecular cloud. Yet, it should be stressed here that the spectral shape gained from modeling the gamma ray emission via neutral pion decay, on the other hand, provides the primary proton spectrum at the location of the formation of the neutral pions, the molecular cloud, since they decay in less than $10^{-16}$~s \citep{particledatabook2010}. The gamma rays emitted from the decay of these pions do hardly suffer energy losses. Therefore, the primary proton spectrum at the location of the cloud is obtained. Low energy protons are very likely accelerated in the same place as the high energy protons, so in this cloud there is not only the formation of pions, but also ionization to be expected. Furthermore, deceleration of high energy protons additionally increases the number of low energy protons \citep{padovani2009}. Since the spectrum obtained from modeling holds for the location of the molecular cloud, no additional momentum losses have to be considered.
However, the primary proton SED can only be considered as known from
the modeling down to energies of roughly 1~GeV, as mentioned
above. Below this energy, there is no observational information about
the primary proton spectrum available. Ionization, on the other hand,
will only be effective below 1~GeV, with the cross section for the
direct ionization of molecular hydrogen by an incoming proton peaking
at about 100~keV and rapidly declining with increasing proton energy
(\cite{padovani2009} and references therein). This makes estimates for
the primary proton spectrum below $\sim$~1~GeV rather uncertain and a
crucial part of the calculation of the ionization rate. It is well
possible and most probable that the power law behavior of the primary
proton SED is not to be extrapolated to lower energies, because the
acceleration mechanisms in these energy domains may differ. In order
to account for loss processes which are effective for energies below
$\sim$~1~GeV as well as the unknown acceleration mechanism at these
energies, here the primary proton spectrum derived from modeling is
attenuated to lower energies by the choice of a broken power law with
positive spectral index $1~\leq~a~\leq~2$, $E^{+a}$, which is
compatible with predictions of \cite{skilling1976}, for three
different lower break energies $E_{\rm lb}$. It should be mentioned
that \cite{zirakashvili2008} and \cite{ellison2011} predict a concave
spectrum for the primary particles escaping SNRs and entering nearby
molecular clouds. The models used there apply for young SNRs with
an age of $10^{3}$~yrs or younger, while here all examined SNRs are at
least middle-aged, about $10^{4}$~yrs or older. For old SNRs, \cite{bozhokin1994} modeled the penetration of a broad cosmic ray proton spectrum into a molecular cloud, assuming a diffusion coefficient with momentum dependence $\kappa~\propto~p^{0.33}$ and $\kappa~\propto~p^{0.5}$. Their results motivate further examination of SNRs interacting with a molecular cloud, as e.g.\ \cite{bykov2000} modeled the spectrum of cosmic ray electrons interacting with a molecular cloud for the case of IC443, where they also derived a profile of the ionization rate due to primary electrons in the shocked part of the cloud. Recently, \cite{yan2011} investigated the acceleration of protons in SNRs and the generation of gamma rays in nearby molecular clouds, taking into account the streaming instability and background turbulence.

\subsection{Acceleration mechanism\label{acc_mech}}

  Fermi
acceleration \citep{fermi1949} is the very first approach to
stochastically accelerate charged particles at shock fronts and the model has been further developed to the theory of diffusive shock
acceleration (DSA) \citep{krymskii77,bell1978,bell1978b,blandford1978,schlickeiser1989a,schlickeiser1989b}. A variety of concrete models concerning the acceleration mechanism at work in
SNRs have been established in the past (e.g.~\cite{scott1975},
\cite{blandford1987}, \cite{blasi2005}, \cite{zirakashvili2010},
\cite{eichler2011}, \cite{drury2011} and references therein). Recently, \cite{uchiyama2010} described an alternate model where the focus is not on escaped CR particles, but strong adiabatic compression behind the SNR shock wave in the cloud leads to diffusive shock acceleration of pre-existing cosmic rays. With their model, they are capable of modeling both the observed synchrotron radiation and gamma ray emission from certain SNRs associated with a molecular cloud, in particular W51C, W44 and IC443.
However, there is no direct evidence which process actually is responsible for the acceleration. Furthermore, propagation effects and possibly reacceleration are important. The unknown nature of the actual acceleration mechanism leads to a huge variety in possible CR spectra below $\sim$~1~GeV, which is shown in figure 1 of \cite{indriolo2009}, where several propagated primary spectra are summarized. But since the modeling of the primary particle spectrum fitting the observed gamma ray emission via neutral pion decay does offer the spectral shape of the primaries at the source above 1~GeV, the primary spectra used in this work are independent of the actual acceleration mechanism. The only assumption made is that the major contribution to the GeV gamma ray emission from the sources is caused by neutral pion decay, which shall be tested this way.

\subsection{Calculation of the proton SED normalization\label{norm}}

In this subsection a detailed description of the calculation of the normalization of the proton SED using the observed gamma spectrum is given.
Known is the flux of gamma rays detected at Earth,
\begin{equation}
J_{\gamma} = \frac{{\mathrm{d}}N_{\gamma}}{{\mathrm{d}}E_{\gamma}~{\mathrm{d}}t~{\mathrm{d}}A_{\rm Earth}}~,
\end{equation}
in units of GeV$^{-1}$~s$^{-1}$~cm$^{-2}$. What can be calculated from observations is the normalization of the flux of high energy protons interacting with the cloud and thus forming neutral pions causing the gamma radiation, which is continued to lower energies in order to calculate the ionization rate of the cloud next to the SNR. This flux is needed not at Earth but at the SNR,
\begin{equation}
j_{\rm p} = \frac{{\mathrm{d}}N_{\rm p}}{{\mathrm{d}}E_{\rm p}~{\mathrm{d}}t~{\mathrm{d}}A_{\rm source}} = a_{\rm p} \Phi_{\rm p}(E_{\rm p}) \label{100}
\end{equation}
in units of GeV$^{-1}$~s$^{-1}$~cm$^{-2}$. 
For instance, for W51C, 
\begin{equation}
\Phi_{\rm p}(E_{\rm p}) = \left(\frac{E_{\rm p}}{1\rm{~GeV}}\right)^{-1.5}\left(1+\frac{E_{\rm p}}{15\rm{~GeV}}\right)^{-1.4}~.
\end{equation}
Here, $\Phi_{\rm p}(E_{\rm p})$ is a dimensionless spectral function and $a_{\rm p}$ is the normalization factor in units of GeV$^{-1}$~s$^{-1}$~cm$^{-2}$. 
The latter enters the calculation of the ionization rate. In order to obtain this value, the proton flux at the source is calculated from the observed gamma ray spectrum.

The calculation of the formation of gamma rays via neutral pion decay is described in detail in e.g.\ \cite{kelner2008}. In this paper, the equation for the formation rate of gamma rays in the energy interval ($E_{\gamma}$, $E_{\gamma}~+~{\mathrm{d}}E_{\gamma}$) and a unit volume from the decay of neutral pions is reported as:
\begin{equation}
\begin{aligned}
\Phi_{\gamma}(E_{\gamma}) &\equiv \frac{{\mathrm{d}}N_{\gamma}}{{\mathrm{d}}E_{\gamma}~{\mathrm{d}}V~{\mathrm{d}}t} \\
&= n_{\rm{H}} \int_{E_{\gamma}}^{\infty}{\sigma_{\rm{inel}}(E_{\rm p})
  j_{\rm p}(E_{\rm p}) F_{\gamma}\left(\frac{E_{\gamma}}{E_{\rm p}},~E_{\rm p}
  \right)~\frac{{\mathrm{d}}E_{\rm p}}{E_{\rm p}}}~,
\label{gamma}
\end{aligned}
\end{equation}
where $n_{\rm{H}}$ is the density of the ambient medium and $\sigma_{\rm{inel}}(E_{\rm p})$ is the cross section of inelastic proton-proton interactions\footnote{\cite{kelner2008} use the CR density, while here the CR flux is used.}. The function $F_{\gamma}(x,~E_{\rm p})$ describes the number of photons in the energy interval ($x, x~+~{\mathrm{d}}x$) per collision and is a dimensionless probability density distribution function.

The result of the formula mentioned above is the number of gamma rays formed from neutral pion decay per unit time, unit energy and unit volume at the location of the hadronic interactions, in units of GeV$^{-1}$~s$^{-1}$~cm$^{-3}$. To transform this quantity into the quantity detected, the result first has to be multiplied by the volume of the interaction region in order to obtain the total number of gamma rays formed from neutral pion decay per unit time and unit energy:
\begin{equation}
\begin{aligned}
\frac{{\mathrm{d}}N_{\gamma}}{{\mathrm{d}}E_{\gamma}~dt} = \Phi_{\gamma}(E_{\gamma}) V_{\rm cloud} = n_{\rm{H}} V_{\rm cloud} \\ \cdot \int_{E_{\gamma}}^{\infty}{\sigma_{\rm{inel}}(E_{\rm p}) j_{\rm p}(E_{\rm p}) F_{\gamma}\left(\frac{E_{\gamma}}{E_{\rm p}},~E_{\rm p} \right)~\frac{{\mathrm{d}}E_{\rm p}}{E_{\rm p}}}~.
\end{aligned}
\end{equation}
Assuming that these gamma rays are emitted isotropically, the fraction that is detected at Earth per unit area can be gained from this by dividing by (4$\pi d^2_{\rm Earth-source}$) to account for the solid angle:
\begin{equation}
J_{\gamma} = \Phi_{\gamma}(E_{\gamma}) V_{\rm cloud} \left(4\pi d^2_{\rm Earth-source}\right)^{-1}~,
\label{101}
\end{equation}

Using eqs.\ (\ref{100}), (\ref{gamma}) and (\ref{101}), the photon
spectrum at Earth can be written as
\begin{equation}
J_{\gamma} = a_{\gamma} \int_{E_{\gamma}}^{\infty}{\sigma_{\rm{inel}}(E_{\rm p}) \Phi_{\rm p}(E_{\rm p}) F_{\gamma}\left(\frac{E_{\gamma}}{E_{\rm p}},~E_{\rm p} \right)~\frac{{\mathrm{d}}E_{\rm p}}{E_{\rm p}}}~, \label{ap-formula}
\end{equation}
where 
\begin{equation}
a_{\gamma} = \frac{a_{\rm p} n_{\rm{H}} V_{\rm cloud}}{4\pi d^2_{\rm Earth-source}} = {\mathrm{const.}} \label{104}
\end{equation}
is the normalization constant of the gamma spectrum. This factor is
determined by high-energy gamma observations of the sources.

For the calculation of ionization rates, the proton SED normalization, $a_{\rm p}$ has to be calculated. This is done using conservation of energy:
\begin{equation}
\begin{aligned}
\int_{E_{\min}}^{\infty}{\frac{{\mathrm{d}}N_{\rm p}}{{\mathrm{d}}E_{\rm p}~{\mathrm{d}}t~{\mathrm{d}}A_{\rm source}}E_{\rm p}~{\mathrm{d}}E_{\rm p}} &= a_{\rm p} \int_{E_{\min}}^{\infty}{\frac{\Phi(E_{\rm p})~E_{\rm p}~{\mathrm{d}}E_{\rm p}}{v(E_{\rm p})}} \\ &= \frac{W_{\rm p}}{V_{\rm cloud}}~,
\end{aligned}
\end{equation}
where $v(E_{\rm p}) = \left(1 - \left(1 + \frac{E_{\rm p}}{m_{\rm p} c^2}\right)^{-2}\right)^{1/2} c$ is the velocity of the particle depending on its kinetic energy $E_{\rm p}$ and $W_{\rm p}$ is the total proton energy budget of protons with a minimum energy of $E_{\min}$. Solving this for $a_{\rm p}$ gives:
\begin{equation}
a_{\rm p} = \frac{W_{\rm p}}{V_{\rm cloud}} \left(\int_{E_{\min}}^{\infty}{\frac{\Phi(E_{\rm p})~E_{\rm p}~{\mathrm{d}}E_{\rm p}}{v(E_{\rm p})}} \right)^{-1}~, \label{103}
\end{equation}
where $a_{\rm p}$ is in units of GeV$^{-1}$~s$^{-1}$~cm$^{-2}$. So the entire expression for the normalization of the gamma spectrum from neutral pion decay, $a_{\gamma}$, can be written as
\begin{equation}
a_{\gamma} = \frac{W_{\rm p} n_{\rm{H}}}{4\pi d^2_{\rm Earth-source}} \left(\int_{E_{\min}}^{\infty}{\frac{\Phi(E_{\rm p})~E_{\rm p}~{\mathrm{d}}E_{\rm p}}{v(E_{\rm p})}} \right)^{-1} = {\mathrm{const.}} \label{102}
\end{equation}
The gamma normalization is therefore independent of the cloud volume, while the proton SED normalization, $a_{\rm p}$, depends on this volume.

The modeling of the gamma rays is done using the Kamae model \citep{kamae2006,karlsson2008} in order to obtain the value of $a_{\gamma}$. Here, a factor of 1.85 is multiplied to the resulting spectrum in order to take helium and heavier nuclei into account, as suggested by \cite{mori2009}. Here, the model is modified by these two factors and will be referred to as the modified Kamae model.

For a given spectral shape of the proton SED, $\Phi_{\rm p}(E_{\rm p})$, the gamma ray spectrum normalization, $a_{\gamma}$, can be found from modeling. For a given distance of the object from Earth, $d_{\rm Earth-source}$, and lower integration threshold, $E_{\min}$, this leads to a certain value for the product $W_{\rm p} \cdot n_{\rm{H}}$ for each object. Because there are no precise estimates of the average hydrogen densities of the objects, $n_{\rm{H}}$, here a value of $n_{\rm{H}} = 100$~cm$^{-3}$ is assumed. The result for $W_{\rm p}$ simply scales inversely with $n_{\rm{H}}$, if $n_{\rm{H}}$ should turn out to be different. With a value for $W_{\rm p}$, the proton SED normalization $a_{\rm p}$ is calculated for each object. Since $a_{\rm p}$ is already implicitly including the solid angle interval, the multiplication by $4\pi$ in equation~(\ref{eq_ion}) is not performed.

\section{Calculation of the ionization rate\label{ion_rate}}

In general, ionization by particles is mainly caused by two different kinds of particles, namely electrons and protons.
One has to consider direct ionization by electrons as well as protons on the one hand and ionization by electron capture
of protons on the other hand.
The full expression for the ionization rate following \cite{padovani2009} is
\begin{equation}
\begin{aligned}
\zeta^{\rm{H_2}} = 4\pi \sum_k \int_{E_{\rm{min}}}^{E_{\max}} j_k(E_k)[1 + \phi_k(E_k)]\sigma_{k}^{\rm ion}(E_k){\mathrm{d}}E_k \\ + 4\pi \int_{0}^{E_{\max}}j_{\rm p}(E_{\rm p})\sigma_{\rm p}^{\rm e.c.}(E_{\rm p}){\mathrm{d}}E_{\rm p}~,
\end{aligned}
\end{equation}

where $E_{\min}$ is the minimum energy of particles considered to contribute to the ionization process, $j_k$ is the number of CR particles of species $k$ ($k = {\rm e}$ or ${\rm p}$, respectively) per unit time, area, solid angle and energy interval, $\sigma_{k}^{\rm ion}$ is the ionization cross section for particles of species $k$, $\sigma_{\rm p}^{\rm e.c.}$ is the electron capture cross section for protons and $\phi_k$ is a number taking into account that ionization may not only be due to ionization by a primary particle $k$, but also by the electrons set free during this ionization, called secondary ionization.
A closer look at the orders of magnitude for the different summands shows that only the contribution from primary proton ionization
is significant at the considered energies. The other ionization processes by protons have cross sections which are significantly lower than the one for direct ionization by primary protons, as can be seen in figure 1 of \cite{padovani2009}.

Ionization by primary electrons is neglected here for two reasons: First, the ionization cross sections for these processes are lower than the corresponding ones for protons. Second, while protons hardly lose energy on their way from the SNR to the molecular cloud, electrons can suffer energy losses. However, since the primary CR spectra near the peak of the corresponding ionization cross section ($\sim 10^5$~eV) is unknown, it is possible that primary electrons do contribute to the ionization rate significantly, as discussed in \cite{padovani2009}.  Because the focus of this work is on the contribution of CR protons, here it is assumed that the contribution of primary electrons to the total ionization rate is dominated by the contribution of primary protons and secondary electrons. In fact, ionization by secondary electrons is an important aspect, but it only increases the ionization rate by less than a factor of 2. The resulting ionization rates derived here are rather lower limits, due to neglecting the contribution of primary electrons.

Since only primary particles with a minimum kinetic energy of $10^5$~eV can penetrate the cloud (as will be discussed below), this contribution seems negligible. On the other hand, protons penetrating the cloud will lose energy gradually, so in principle the effect of secondary ionization needs to be calculated differentially, as was done in \cite{padovani2009}. This will be done in future work and thus the ionization rates calculated here are lower limits. 

Neglecting the aforementioned ionization processes, the calculation reduces to:
\begin{equation}
\zeta^{\rm{H_2}} = 4\pi \int_{E_{\min}}^{E_{\max}} j_{\rm p}(E_{\rm p})\sigma_{\rm p}^{\rm ion}(E_{\rm p}){\mathrm{d}}E_{\rm p}~. \label{eq_ion}
\end{equation}
For the examined SNRs the lower limit of integration, $E_{\min}$, depends on the hydrogen density. This is due to the fact that the CR particles have to penetrate the molecular cloud before interacting with the gas, and lower energy particles are decelerated on shorter length scales than higher energetic ones and therefore do not contribute noticeably to the ionization. However, deceleration of high energy particles populates the low energy part of the spectrum. The stopping length of the particles depends on the hydrogen density, and thus does the lower integration limit. \cite{indriolo2009} suggest a lower integration threshold of 2~MeV for diffuse clouds as originally suggested by \cite{spitzer1968}, $n_{\rm{H}} < 10^3$~cm$^{-3}$, and a lower integration threshold of 10~MeV for dense clouds, $n_{\rm{H}} > 10^3$~cm$^{-3}$. But since this effect is not well understood in quantitative terms, 100~MeV for dense clouds or 100~keV for diffuse clouds in the direct vicinity of the SNR might be considered as well.
The direct ionization cross section for primary protons used is reported by \cite{padovani2009} in eqs. (5) and (6). 

Using the proton spectra obtained from modeling the gamma ray detections, the ionization rate by primary protons is calculated from equation~(\ref{eq_ion}). These spectra are varied in the spectral index $a$ below the lower break energy $E_{\rm lb}$. Additionally, the lower break energy $E_{\rm lb}$ is varied. The minimum energy of protons penetrating the cloud and thus contributing to the ionization process considered is $E_{\min}$~$=$~$10$~MeV as a conservative approximation. The Galactic average ionization rate of molecular hydrogen in molecular clouds is $\zeta^{\rm{H_2}}_{\rm{gal.~aver.}}$~$=$~$2$~$\times 10^{-16}$~s$^{-1}$, as reported by \cite{neufeld2010}. It should be noted that this average value is subject to variations between approximately $10.5$~$\times 10^{-16}$~s$^{-1}$ at maximum and $0.5$~$\times 10^{-16}$~s$^{-1}$ or less, possibly connected to propagation effects \citep{indriolo2012}. 
The proton spectra of the sources are either of the form

\begin{equation}
j_{\rm p}(E_{\rm p}) = \left\{\begin{array}{cl}
  a_{\rm p}\left(\frac{E_{\rm lb}}{E_0}\right)^{-s}\left(1+\frac{E_{\rm lb}}{E_{\rm br}}\right)^{-\Delta s} \left(\frac{E_{\rm p}}{E_{\rm lb}}\right)^{a} & (E_{\rm p} \leq E_{\rm lb}) \\
  a_{\rm p}\left(\frac{E_{\rm p}}{E_0}\right)^{-s}\left(1+\frac{E_{\rm p}}{E_{\rm br}}\right)^{-\Delta s}~ & (E_{\rm p} > E_{\rm lb}),\end{array}\right. 
\end{equation}
or of the form

\begin{equation}
j_{\rm p}(E_{\rm p}) = \left\{\begin{array}{cl}
  a_{\rm p}\left(\frac{E_{\rm lb}}{E_0}\right)^{-s} \exp{\left(-\frac{E_{\rm lb}}{E_{\rm cutoff}}\right)} \left(\frac{E_{\rm p}}{E_{\rm lb}}\right)^{a} & (E_{\rm p} \leq E_{\rm lb}) \\
  a_{\rm p}\left(\frac{E_{\rm p}}{E_0}\right)^{-s} \exp{\left(-\frac{E_{\rm p}}{E_{\rm cutoff}}\right)} & (E_{\rm p} > E_{\rm lb}),\end{array}\right. 
\end{equation}

where $a_{\rm p}$ is the normalization factor, $E_0$~=~1~GeV or 1~TeV (see Table \ref{parameters}), $E_{\rm br}$ is the location of the spectral break, $s \equiv \alpha_l$ is the lower spectral index, $\Delta s + s \equiv \alpha_h$ is the higher spectral index, $E_{\rm cutoff}$ is the higher cutoff energy, $E_{\rm lb}$ is the location of the lower spectral break and $a = 2.0, 1.5$ or $1.0$  the spectral index below the lower break $E_{\rm lb}$. It should be mentioned that the spectral break in the primary proton spectrum is partly due to the fact that it is given in terms of the kinetic energy $E_{\rm p}$. Diffusive shock acceleration produces a power law in momentum, so expressing this spectral behavior in terms of the kinetic energy $E_{\rm p}$, the spectrum deviates from a power law in kinetic energy near the rest energy of the particle, $\sim$~1~GeV for protons.
Because the particles are accelerated at the supernova shock front
which in some cases penetrates the molecular cloud, both acceleration
and ionization can occur in the same place and at the same time. To account for the unknown acceleration mechanism below $E$~$\sim$~1~GeV, the lower spectral break $E_{\rm lb}$ is introduced. Below this break energy, the particle spectrum is assumed to decrease rapidly toward lower particle energies as to give a conservative lower limit on the ionization rate. For $E_{\rm lb} \leq 1$ GeV, this does not change the resulting gamma ray spectrum from neutral pion decay.

Here, $j_{\rm p}(E_{\rm p}) = {\mathrm{d}}N_{\rm p}/({\mathrm{d}}E_{\rm p}~{\mathrm{d}}t~{\mathrm{d}}A_{\rm source})$ is the number of CR protons per unit time, area and energy at the source. The spectral shape of the proton spectrum $\Phi_{\rm p}(E_{\rm p})$ for each object is found modeling the gamma ray detections by \textit{Fermi}-LAT (see references \cite{abdo(W51C)2009}, \cite{abdo(W44)2010}, \cite{abdo(W28)2010}, \cite{abdo(IC443)2010}, \cite{abdo(W49B)2010}, \cite{castro2010}) assuming hadronic interactions to form neutral pions, which decay via gamma-gamma coincidences to be the cause of the gamma ray emission. The spectral information about all sources is given in Table \ref{parameters}.

\begin{landscape}
\begin{table}
	\centering
	{\tiny
		\begin{tabular}[htbp]{|c|c|c|c|c|c|c|c|}
			\hline
  		object & d (kpc) & $V_{\rm cloud}$ (cm$^{3}$) & $E_0$ & $\alpha_l$ & $\alpha_h$ or $E_{\rm cutoff}$ & $W_{\rm p}$ (erg), in protons of $E$ $>$ 10 MeV & spectral break energy $E_{\rm br}$ (GeV) \\
 		  \hline\hline
  		W51C & 6.0 & $3.3 \times 10^{60}$ & 1 GeV & 1.5 & 2.9 & $7.7 \times 10^{49}$ & 15 \\
  		\hline
  		W44 & 3.0 & $4.2 \times 10^{59}$ & 1 GeV & 1.74 & 3.7 & $1.2 \times 10^{50}$ & 9 \\
  		\hline
  		W28 & 2.0 & $3.2 \times 10^{59}$ & 1 GeV & 1.7 & 2.7 & $3.3 \times 10^{49}$ & 2 \\
  		\hline
  		IC443 & 1.5 & $4.2 \times 10^{59}$ & 1 GeV & 2.0 & 2.87 & $2.4 \times 10^{49}$ & 69 \\
  		\hline
  		W49B & 8.0 & $6.3 \times 10^{56}$ & 1 GeV & 2.0 & 2.7 & $4.4 \times 10^{50}$ & 4 \\
  		\hline
  		G349.7+0.2 & 22.0 & $2.4 \times 10^{59}$ & 1 TeV & 1.7 & 0.16 TeV & $2.2 \times 10^{50}$ & - \\
  		\hline
  		CTB 37A & 11.3 & $3.3 \times 10^{60}$ & 1 TeV & 1.7 & 0.08 TeV & $1.3 \times 10^{50}$ & - \\
  		\hline
  		3C 391 & 8.0 & $3.4 \times 10^{59}$ & 1 TeV & 2.4 & 100 TeV & $2.3 \times 10^{50}$ & - \\
  		\hline
  		G8.7-0.1 & 4.5 & $9.3 \times 10^{59}$ & 1 TeV & 2.45 & 100 TeV & $3.7 \times 10^{50}$ & - \\
  		\hline
		\end{tabular}
		}
\begin{center}		
	\caption{Table of parameters used. All values for $W_{\rm p}$ are calculated for $n_{\rm{H}}~=~100$~cm$^{-3}$. \label{parameters}}
\end{center}	
\end{table}
\end{landscape}

\begin{figure*}
	\centering
		\begin{tabular}{ccc}
			\includegraphics[scale=0.3]{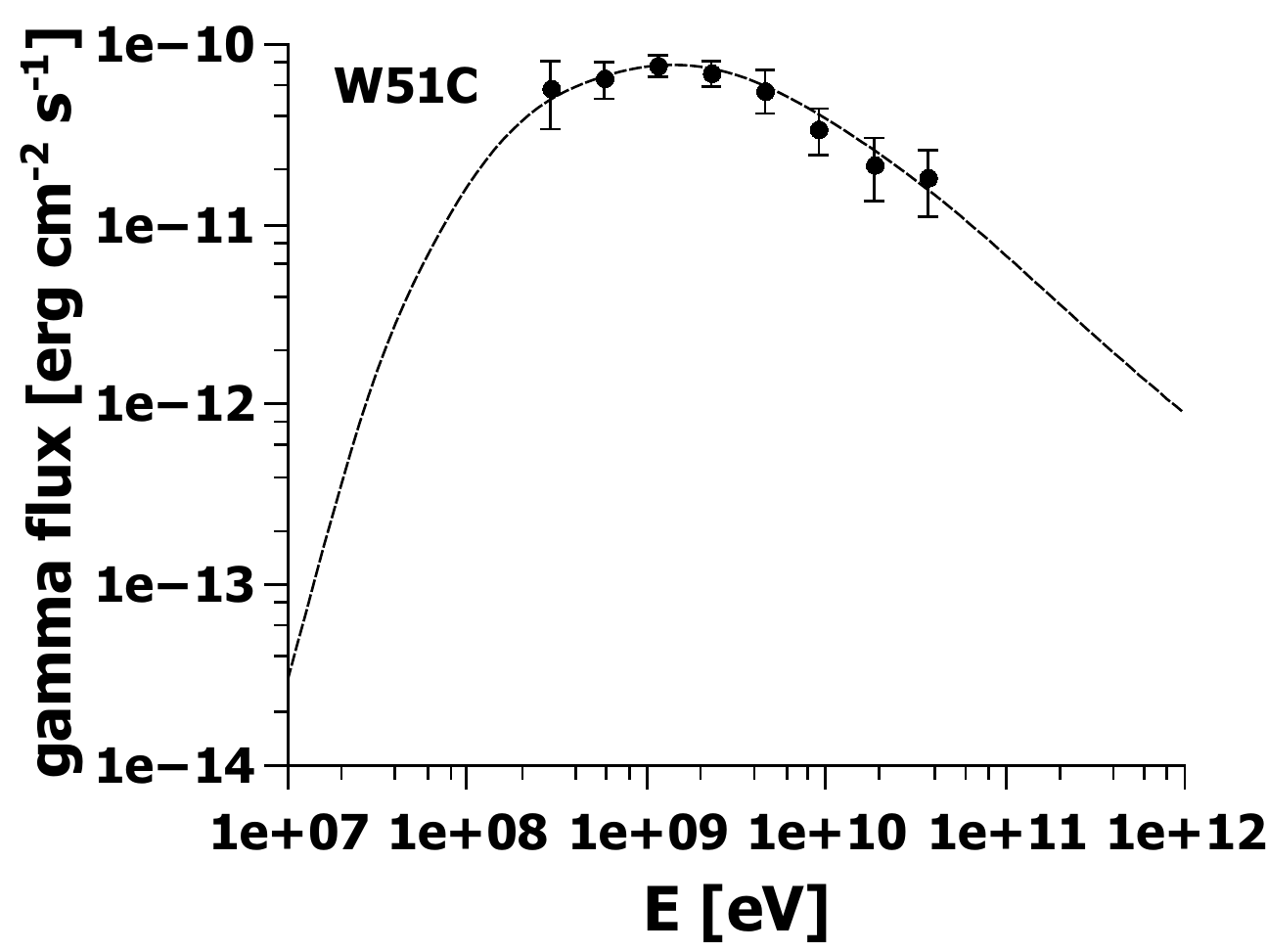} & \includegraphics[scale=0.3]{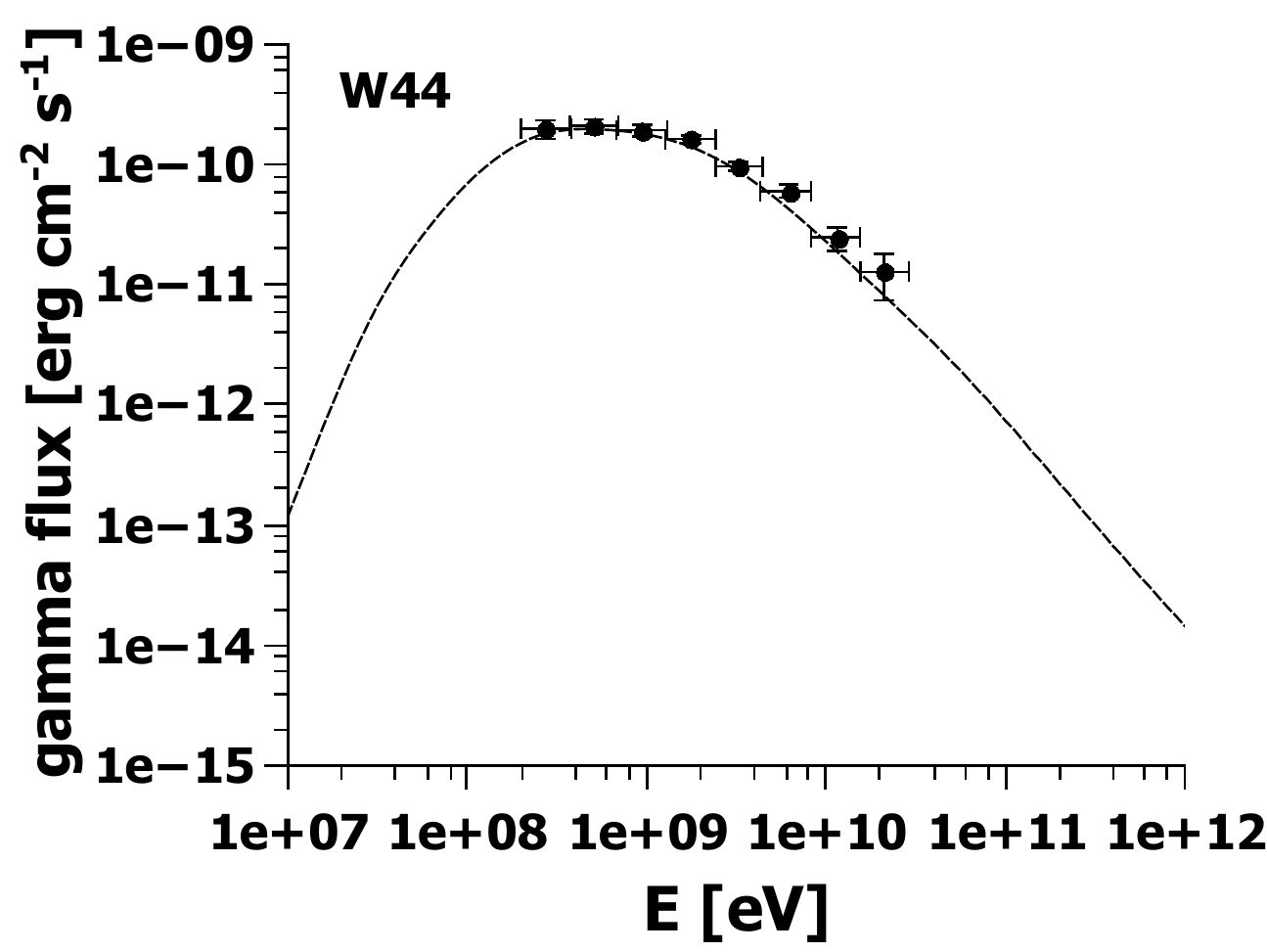} & \includegraphics[scale=0.3]{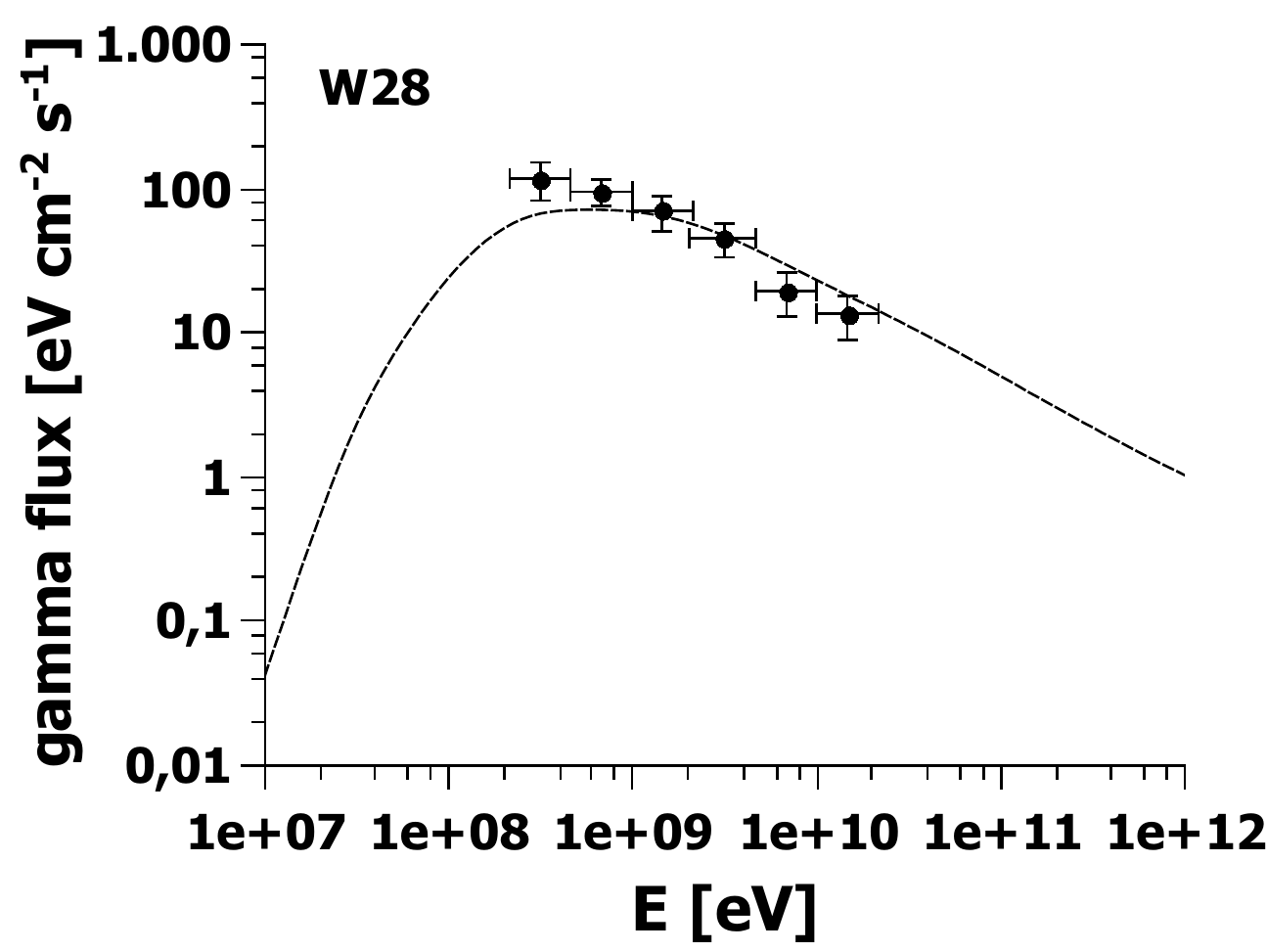} \\
			\includegraphics[scale=0.3]{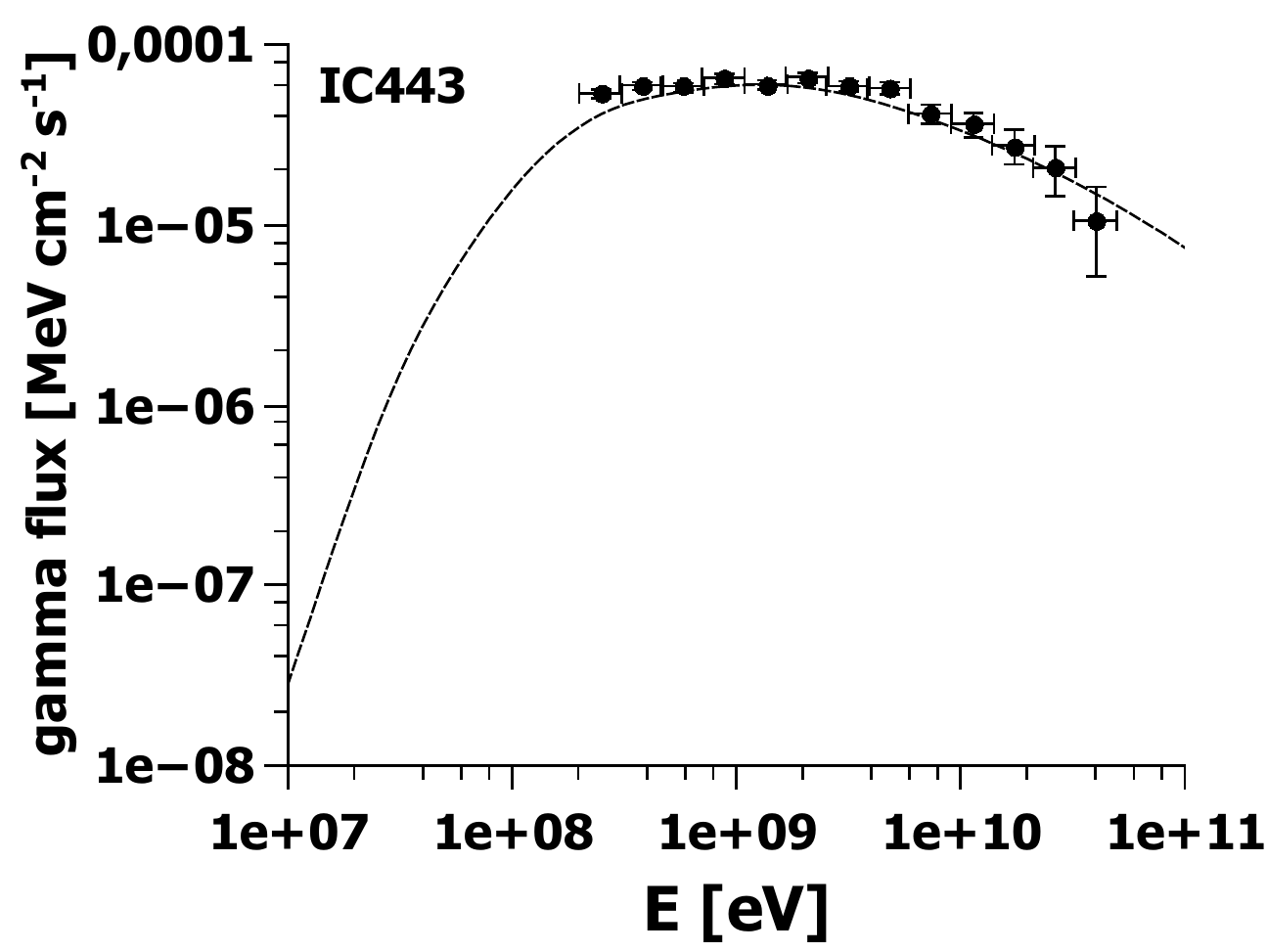} & \includegraphics[scale=0.3]{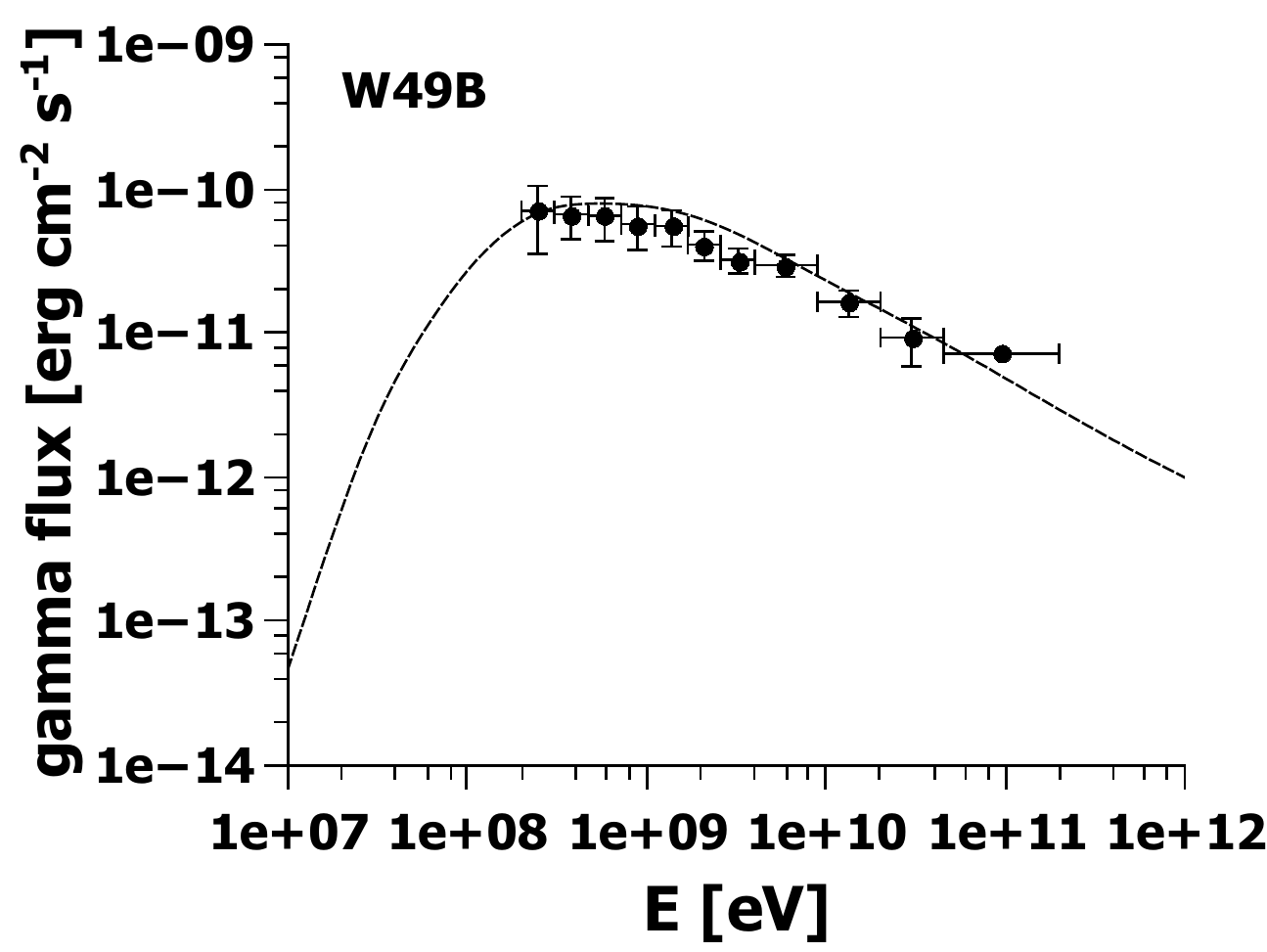} & \includegraphics[scale=0.3]{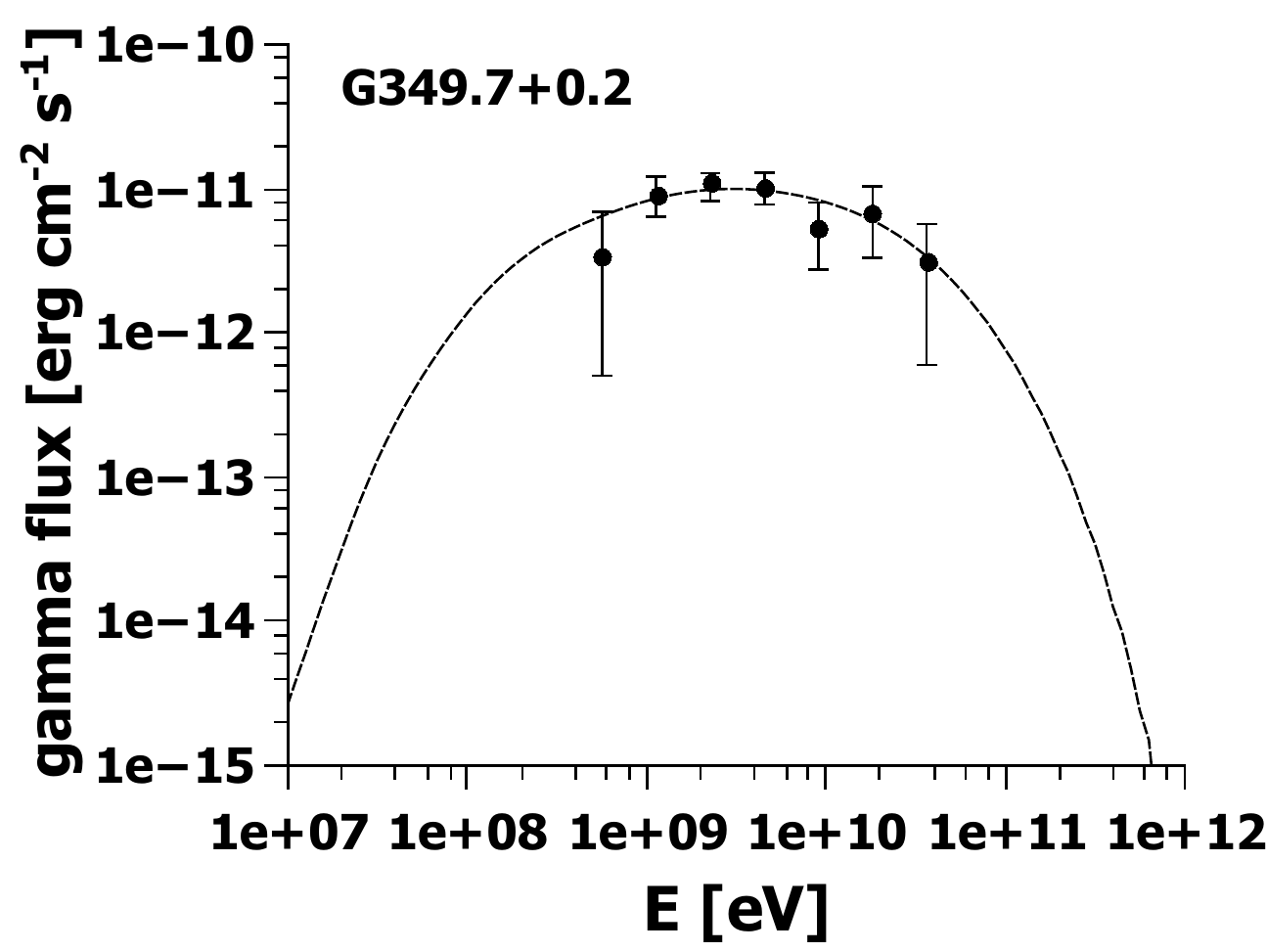} \\
			\includegraphics[scale=0.3]{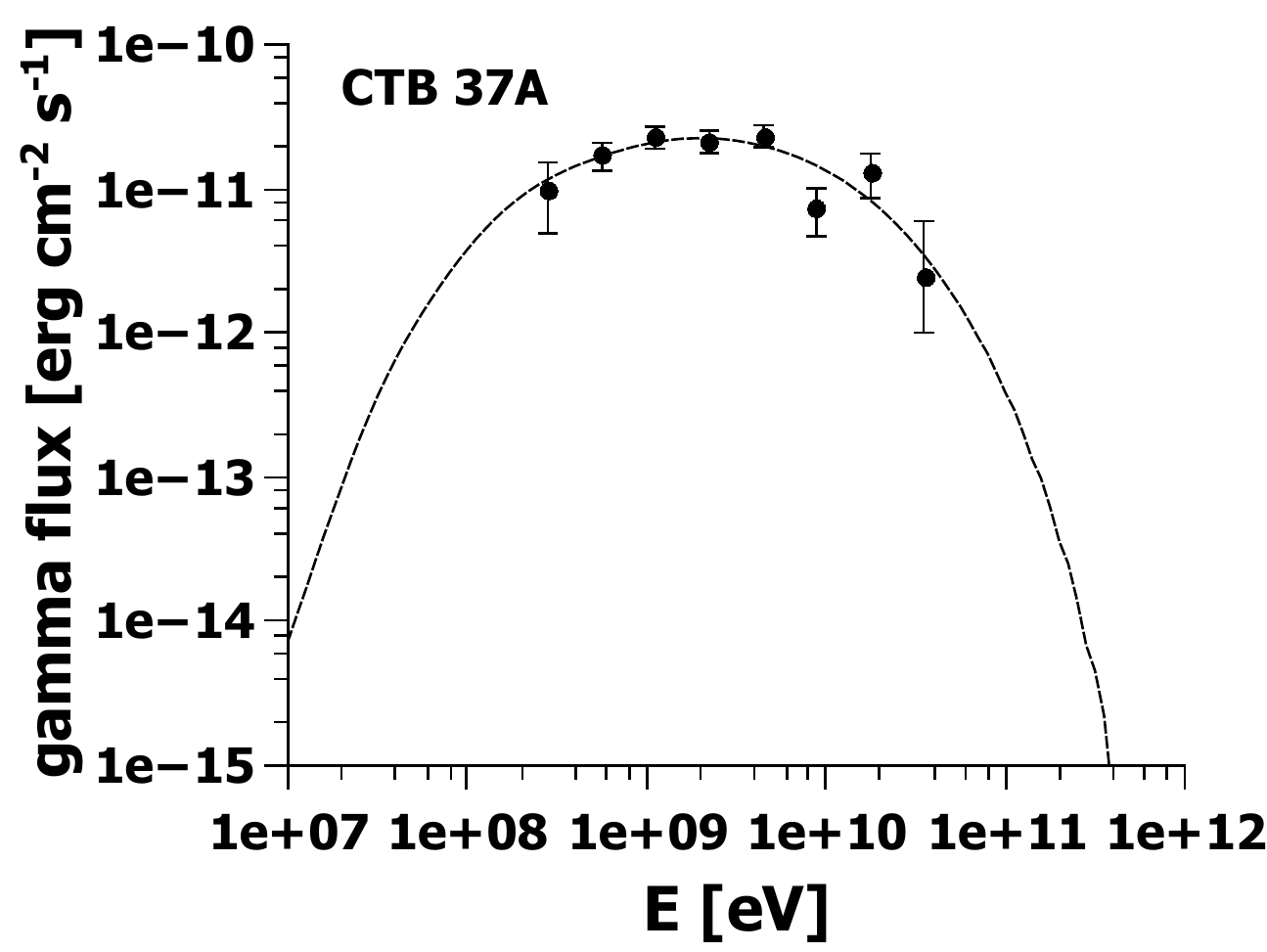} & \includegraphics[scale=0.3]{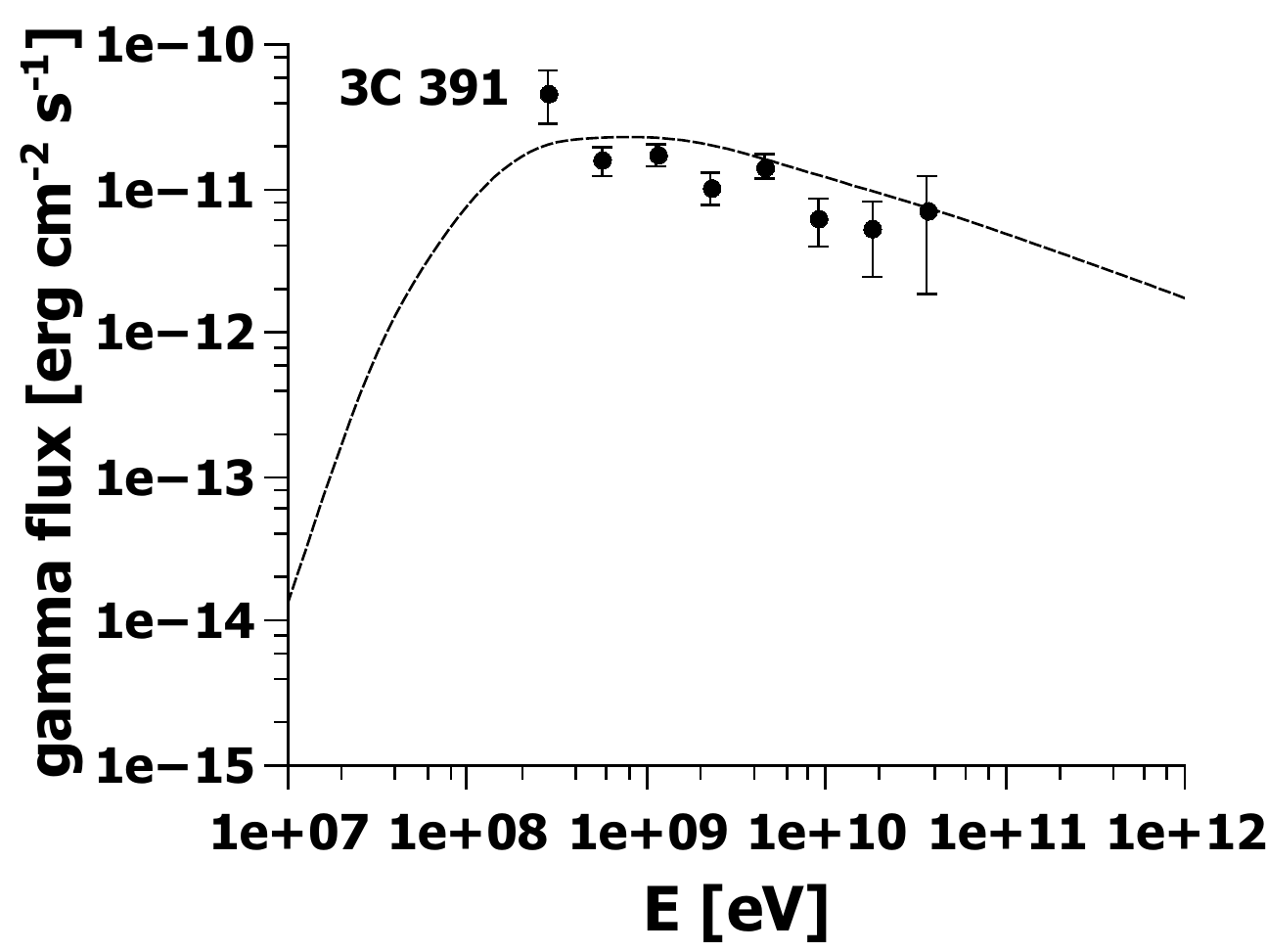} & \includegraphics[scale=0.3]{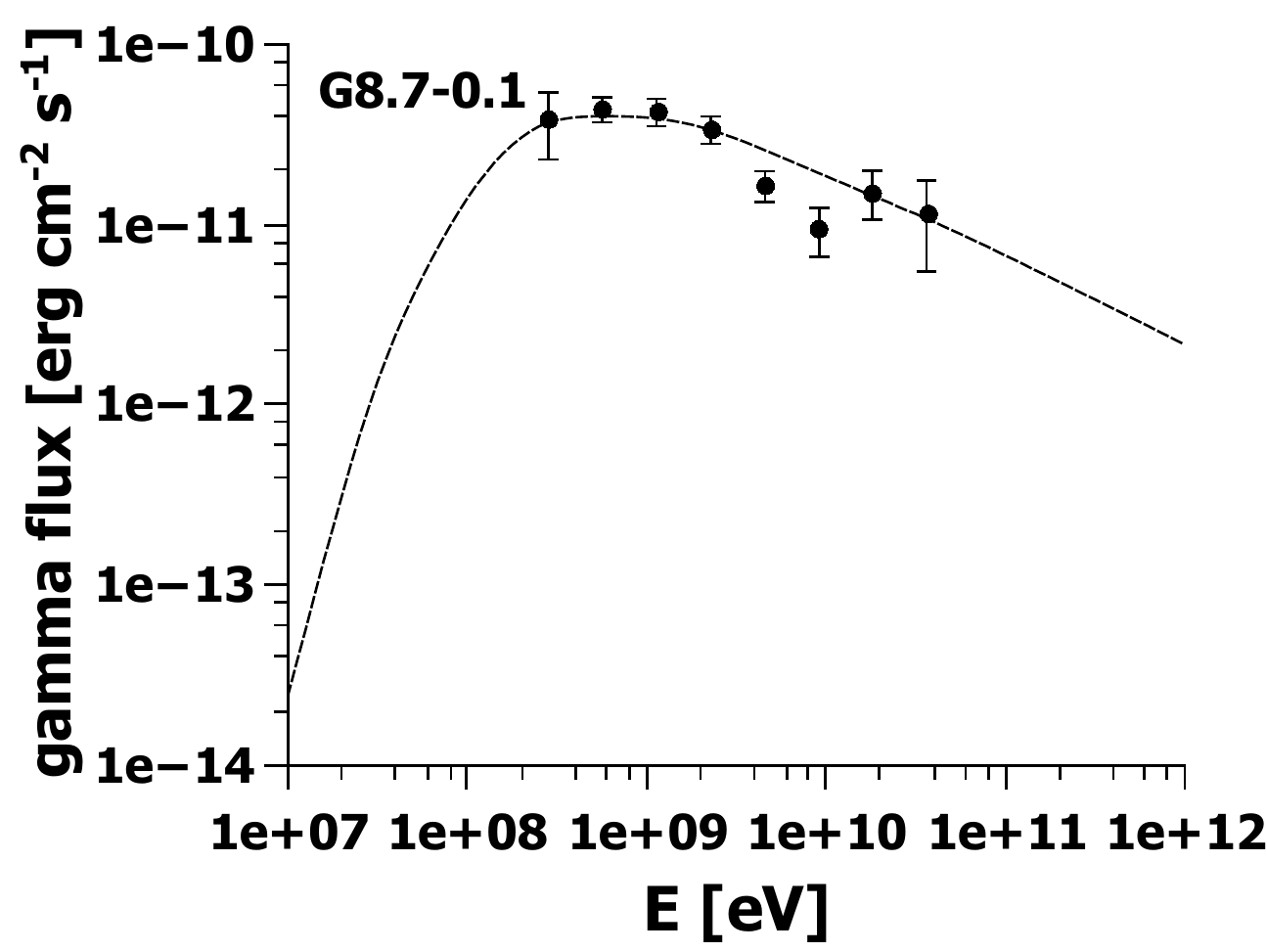} \\
		\end{tabular}
\begin{center}		
	\caption{Modeled gamma ray spectrum and \textit{Fermi}-LAT observational data. The modeled spectrum shown is for $E_{\rm lb}~=~1$~GeV and $a~=~2.0$, but the spectra for $E_{\rm lb}$ down to 30~MeV and $a$ down to 1.0 practically coincide with the spectrum shown due to the low cross section below 1 GeV.}
	\label{gamma_plots}
\end{center}	
\end{figure*}

To calculate the primary proton SED, observational data about the distance $d$ of the object from Earth and the approximate volume of the cloud is required (see equations \ref{ap-formula} and \ref{102}). The values used here are shown in Table \ref{parameters}. The calculation of the ionization rate is discussed here at the example of W51C, but for the other objects the same procedure is done to obtain the result.

In the case of W51C, for $d = 6$~kpc \citep{abdo(W51C)2009}, $E_{\rm lb} = 1$~GeV, $a = 2.0$ and $E_{\min} = 10^7$~eV, only the total proton energy budget of protons with a minimum kinetic energy of $E_{\min}$, $W_{\rm p}$ and the average hydrogen density of the cloud are free parameters (see equation~\ref{102}). The product of these two quantities needs to be 
\begin{equation}
	W_{\rm p} n_{\rm{H}} = 7.7 \times 10^{51} \rm{~erg~cm^{-3}}
\end{equation}
to produce the modeled gamma spectrum shown in Fig.\ \ref{gamma_plots}. As can be seen, the modeled spectrum matches the detections well.

Because there is no precise estimate of the average density of the cloud, here $n_{\rm{H}}$~$=$~$100$~cm$^{-3}$ is assumed, thus offering the required value of $W_{\rm p}$. The result for $W_{\rm p}$ simply scales inversely with $n_{\rm{H}}$, if $n_{\rm{H}}$ should turn out to be different.

The normalization factor of the proton SED, $a_{\rm p}$ is calculated following equation~(\ref{103}),
where $W_{\rm p}$ is the total interacting proton energy budget and the volume of the source $V_{\rm cloud}$ is assumed to be spherical with a radius taken from observations \citep{abdo(W51C)2009,abdo(W44)2010,abdo(W28)2010,abdo(IC443)2010,abdo(W49B)2010,castro2010}. The lower integration limit indicates the minimum energy of particles contributing to ionization processes. As a conservative approximation, here $E_{\min}~=~10$~MeV is used.
The result for each object is shown in Table \ref{tab_p_SED_norm}.

\begin{table*}
\centering{
\begin{tabular}{|c|c|c|c|}
	\hline
  object & $E_{\rm lb}$ & $a_{\rm p}(a = 2.0)$ & $a_{\rm p}(a = 1.0)$ \\
  \hline
  W51C & 1 GeV & $4.0 \times 10^{4}$ & $4.0 \times 10^{4}$ \\
  W51C & 100 MeV & $3.2 \times 10^{4}$ & $3.1 \times 10^{4}$ \\
  W51C & 30 MeV & $2.9 \times 10^{4}$ & $2.9 \times 10^{4}$ \\
  \hline
  W44 & 1 GeV & $1.5 \times 10^{6}$ & $1.4 \times 10^{6}$ \\
  W44 & 100 MeV & $6.9 \times 10^{5}$ & $6.7 \times 10^{5}$ \\
  W44 & 30 MeV & $5.4 \times 10^{5}$ & $5.4 \times 10^{5}$ \\
  \hline
  W28 & 1 GeV & $6.0 \times 10^{5}$ & $5.8 \times 10^{5}$ \\
  W28 & 100 MeV & $2.8 \times 10^{5}$ & $2.7 \times 10^{5}$ \\
  W28 & 30 MeV & $2.1 \times 10^{5}$ & $2.0 \times 10^{5}$ \\
  \hline
  IC443 & 1 GeV & 120 & 120 \\
  IC443 & 100 MeV & 56 & 54 \\
  IC443 & 30 MeV & 34 & 33 \\
  \hline
  W49B & 1 GeV & $3.2 \times 10^{10}$ & $3.1 \times 10^{10}$ \\
  W49B & 100 MeV & $1.3 \times 10^{10}$ & $1.2 \times 10^{10}$ \\
  W49B & 30 MeV & $7.8 \times 10^{8}$ & $7.5 \times 10^{8}$ \\
  \hline
  G349.7+0.2 & 1 GeV & $7.8 \times 10^{-1}$ & $7.7 \times 10^{-1}$ \\
  G349.7+0.2 & 100 MeV & $6.3 \times 10^{-1}$ & $6.3 \times 10^{-1}$ \\
  G349.7+0.2 & 30 MeV & $5.6 \times 10^{-1}$ & $5.6 \times 10^{-1}$ \\
  \hline
  CTB 37A & 1 GeV & $4.4 \times 10^{-1}$ & $4.4 \times 10^{-1}$ \\
  CTB 37A & 100 MeV & $3.4 \times 10^{-1}$ & $3.3 \times 10^{-1}$ \\
  CTB 37A & 30 MeV & $2.9 \times 10^{-1}$ & $2.9 \times 10^{-1}$ \\
  \hline
  3C 391 & 1 GeV & $1.8 \times 10^{-1}$ & $1.7 \times 10^{-1}$ \\
  3C 391 & 100 MeV & $4.8 \times 10^{-2}$ & $4.5 \times 10^{-2}$ \\
  3C 391 & 30 MeV & $2.0 \times 10^{-2}$  & $1.8 \times 10^{-2}$ \\
  \hline
  G8.7-0.1 & 1 GeV & $8.1 \times 10^{-2}$ & $7.8 \times 10^{-2}$ \\
  G8.7-0.1 & 100 MeV & $1.9 \times 10^{-2}$ & $1.8 \times 10^{-2}$ \\
  G8.7-0.1 & 30 MeV & $7.2 \times 10^{-3}$ & $6.8 \times 10^{-3}$ \\
  \hline
\end{tabular}
\caption{Proton SED normalization $a_{\rm p}$ for each source for different lower break energies $E_{\rm lb}$ and spectral indices $a$ below the lower break energy in [erg$^{-1}$ s$^{-1}$ cm$^{-2}$]. \label{tab_p_SED_norm}}
}
\end{table*}

With this normalization one can compute the ionization rate due to primary proton ionization performing the integration~(\ref{eq_ion}) for different values of $E_{\rm lb}$ and a fixed value of $E_{\max}$~=~1~GeV. The lower break energy is of large importance for the result, because the ionization cross section is the largest at low energies and declines rapidly with increasing energy. The upper integration limit may be any value from 1~GeV or higher because of the low ionization cross section for high energies. Changing $E_{\max}$ to 1~PeV does not change the ionization rate significantly. However, changing $E_{\rm lb}$ from 1~GeV to a different value results in a different gamma spectrum normalization $a_{\gamma}$, as can be seen in equation~(\ref{102}).

\begin{figure*}
	\centering
		\begin{tabular}{ccc}
			\includegraphics[scale=0.3]{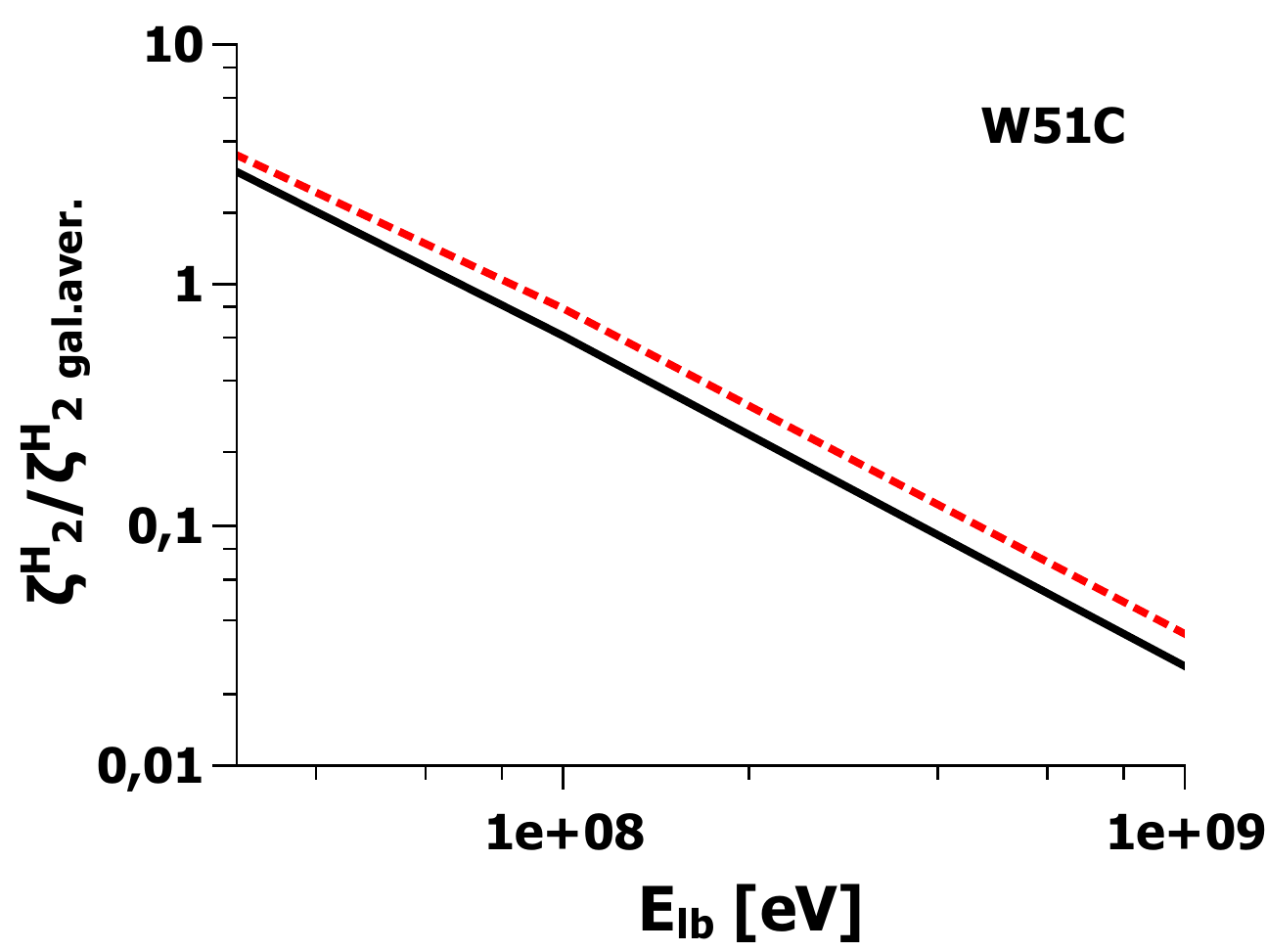} & \includegraphics[scale=0.3]{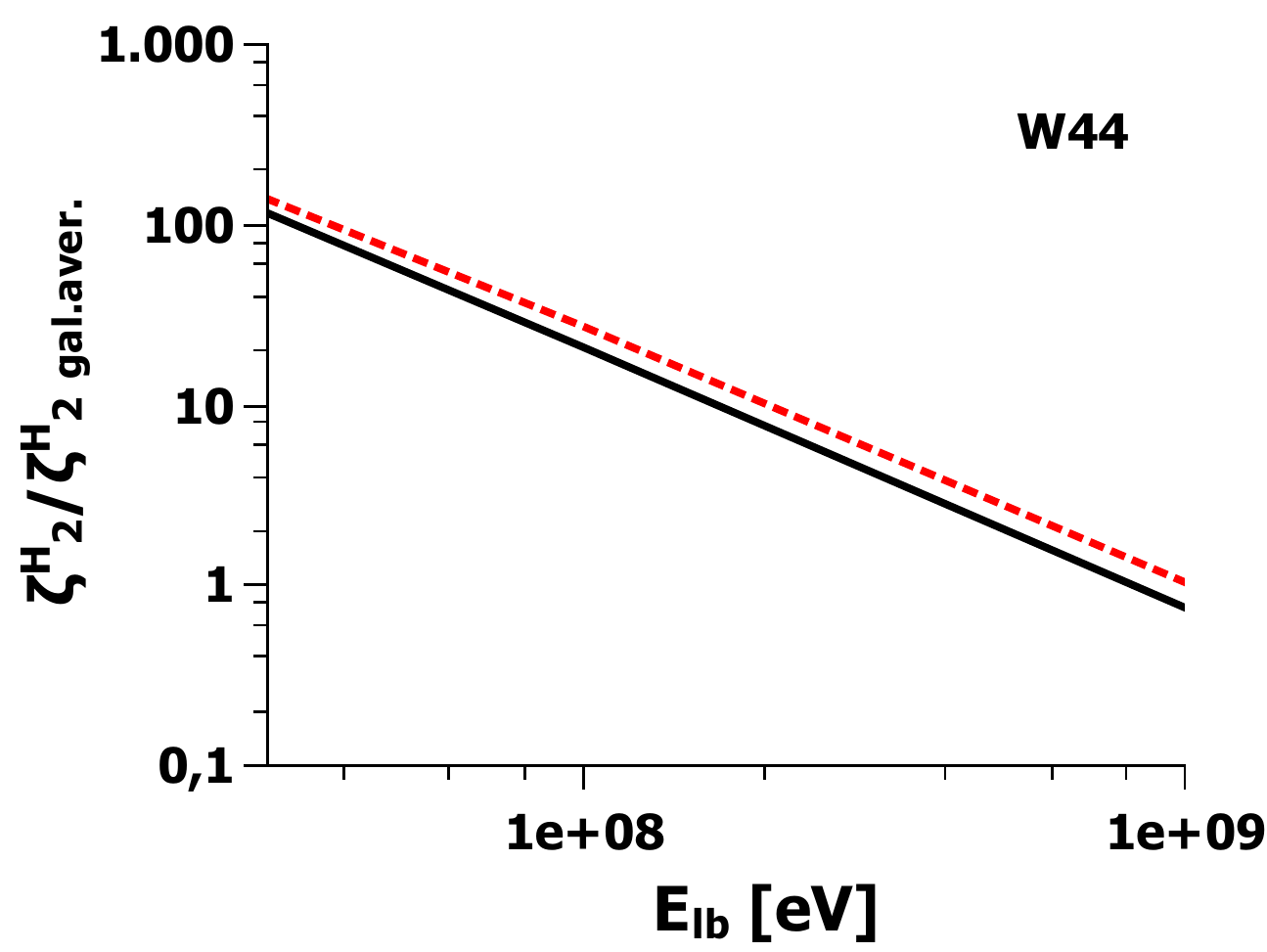} & \includegraphics[scale=0.3]{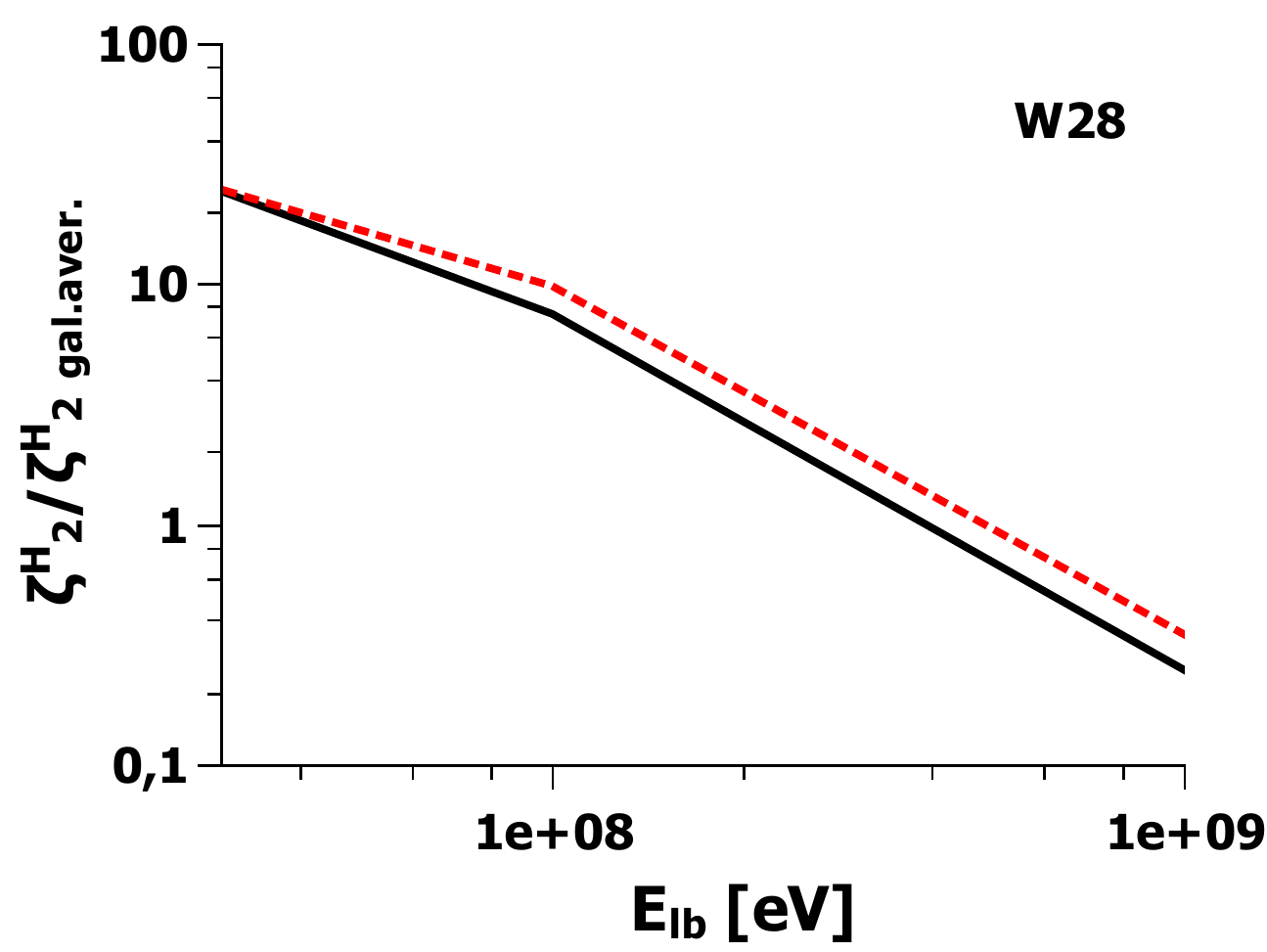} \\
			\includegraphics[scale=0.3]{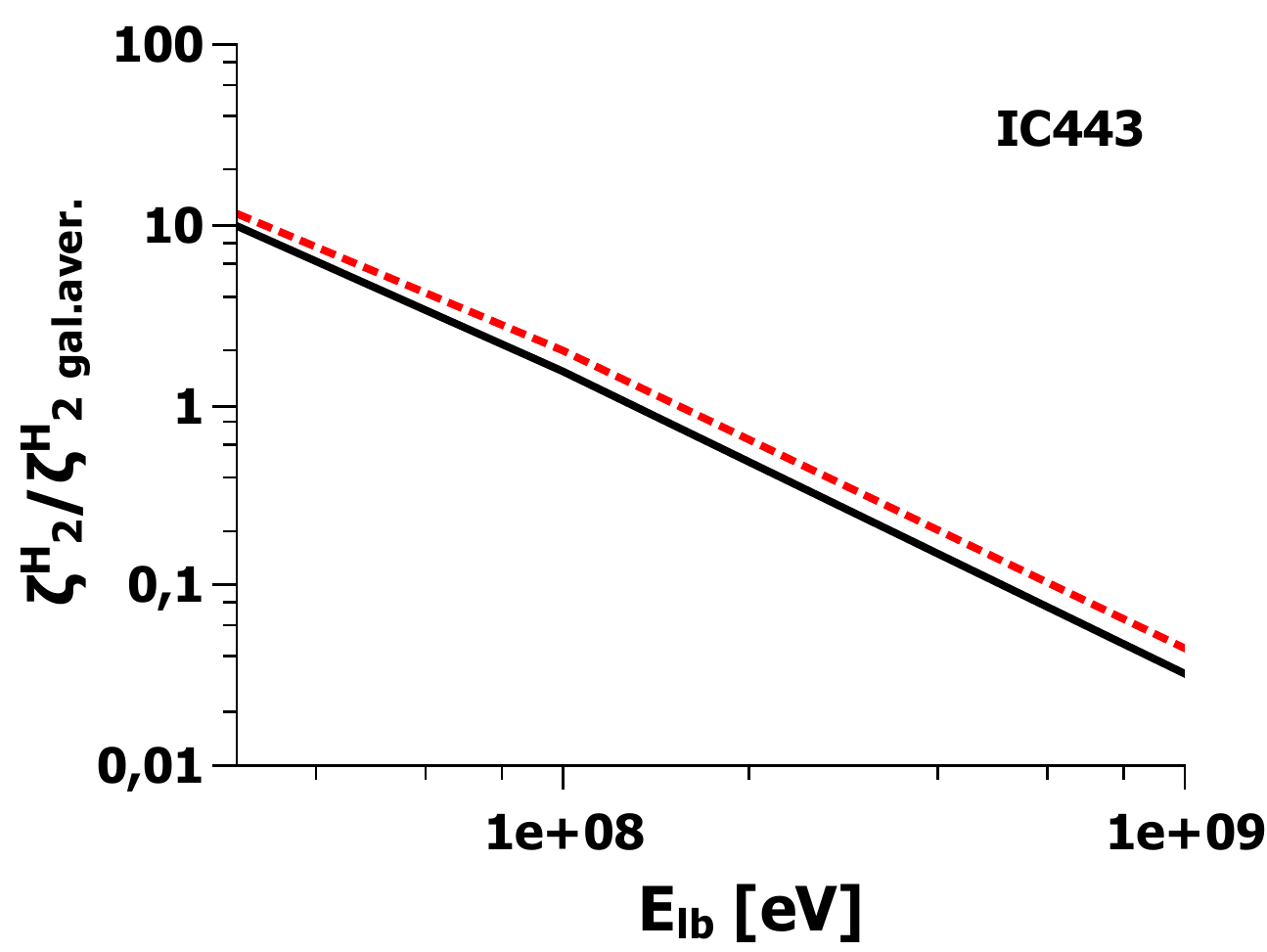} & \includegraphics[scale=0.3]{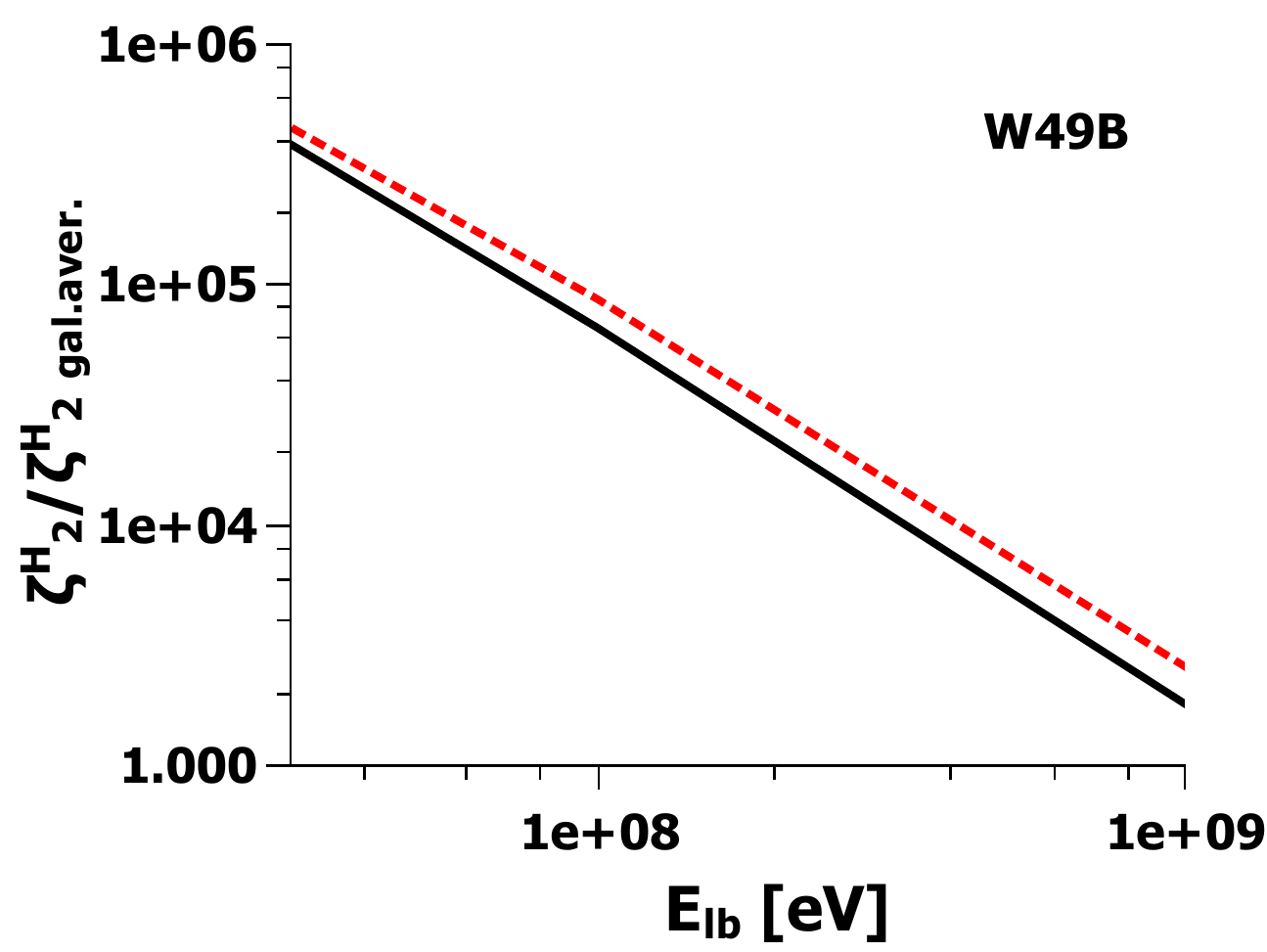} & \includegraphics[scale=0.3]{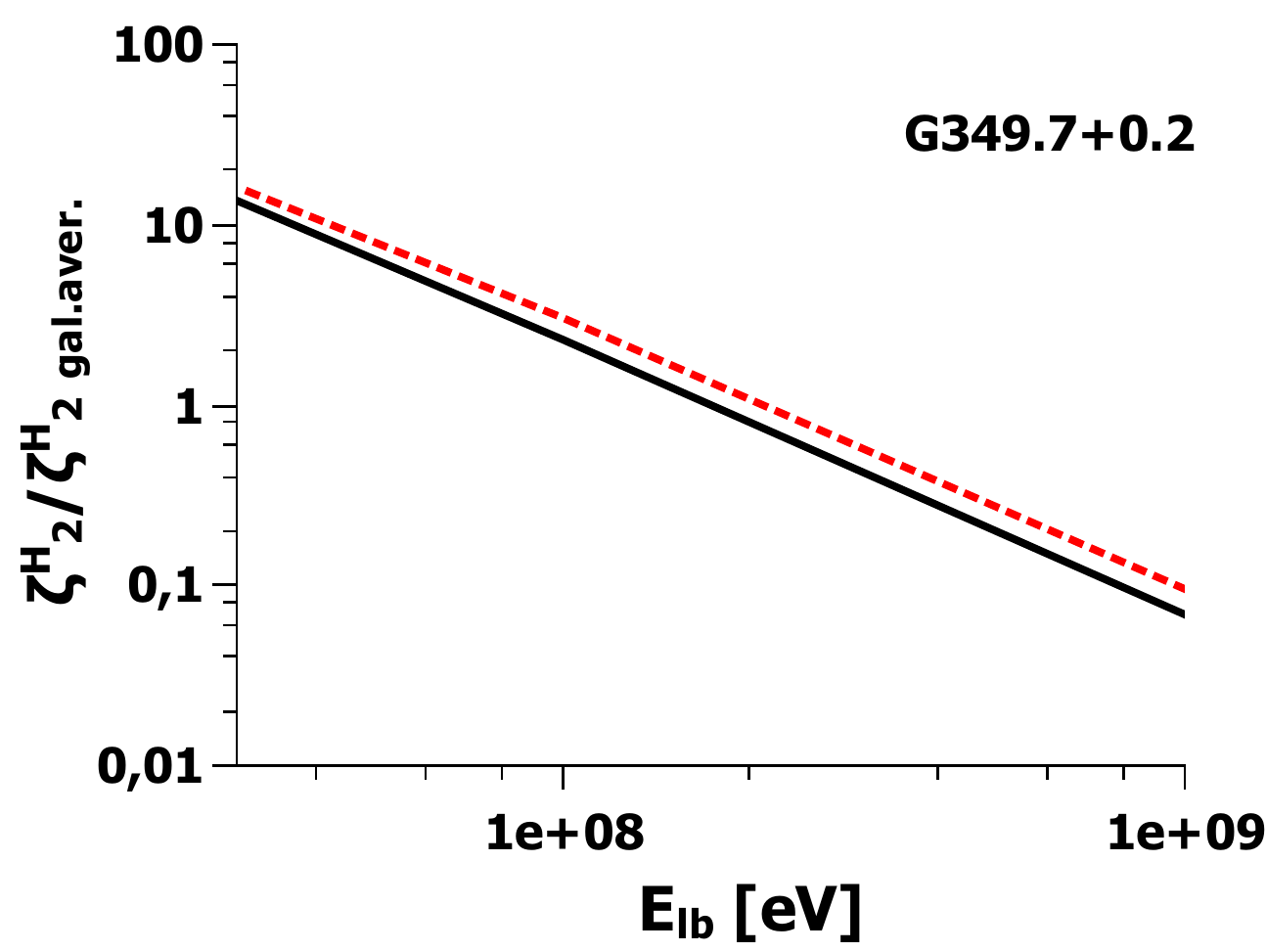} \\
			\includegraphics[scale=0.3]{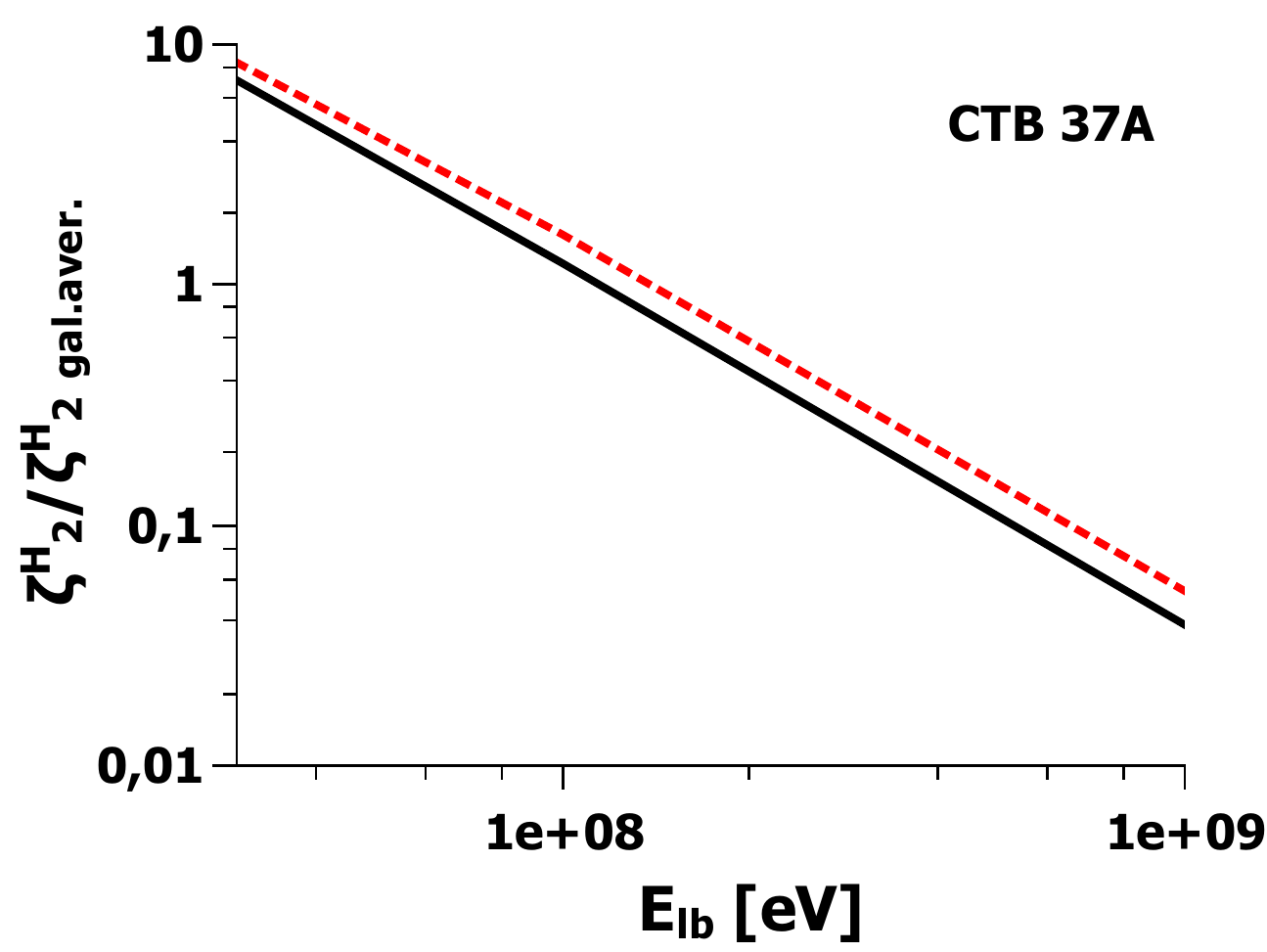} & \includegraphics[scale=0.3]{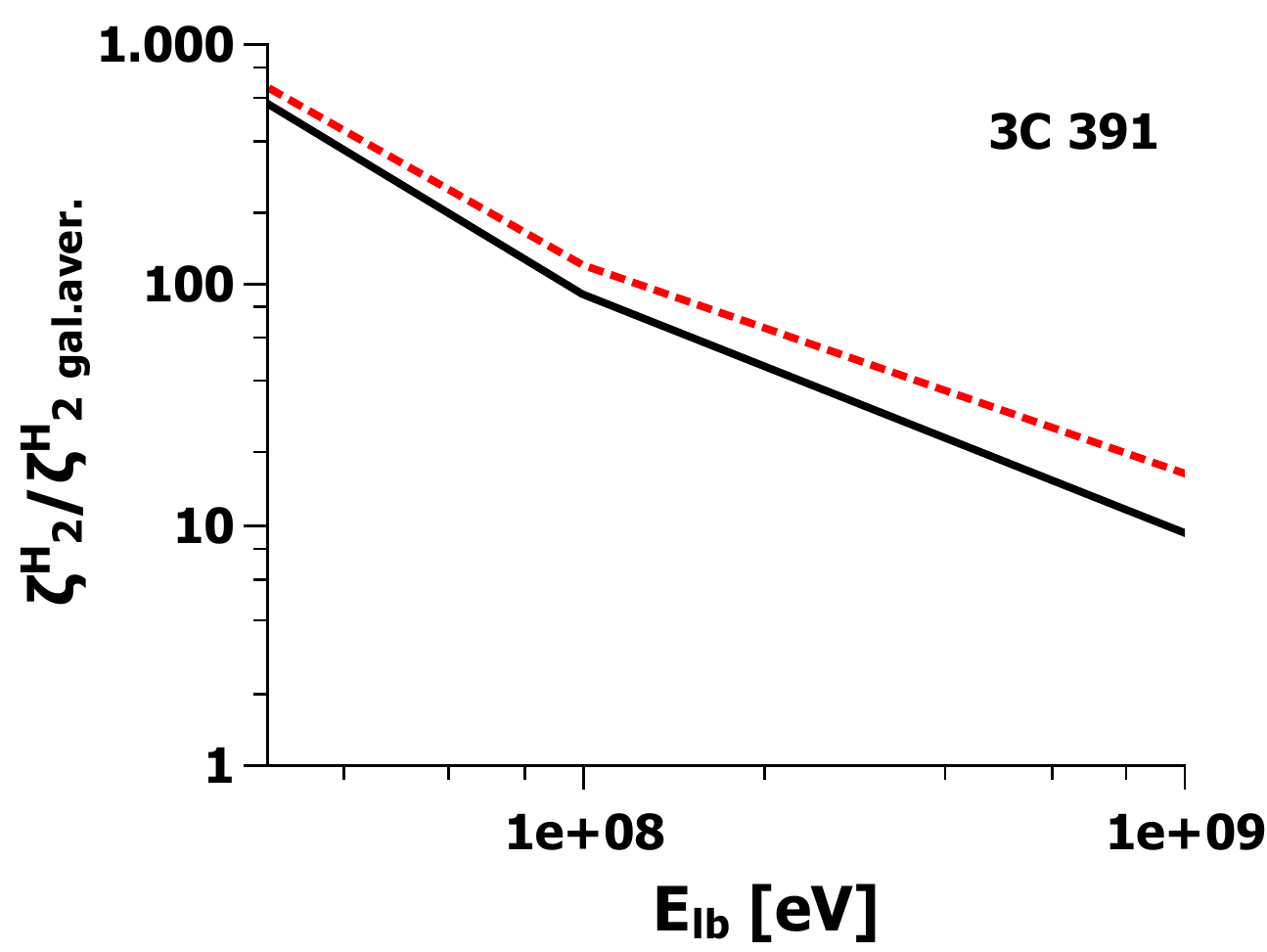} & \includegraphics[scale=0.3]{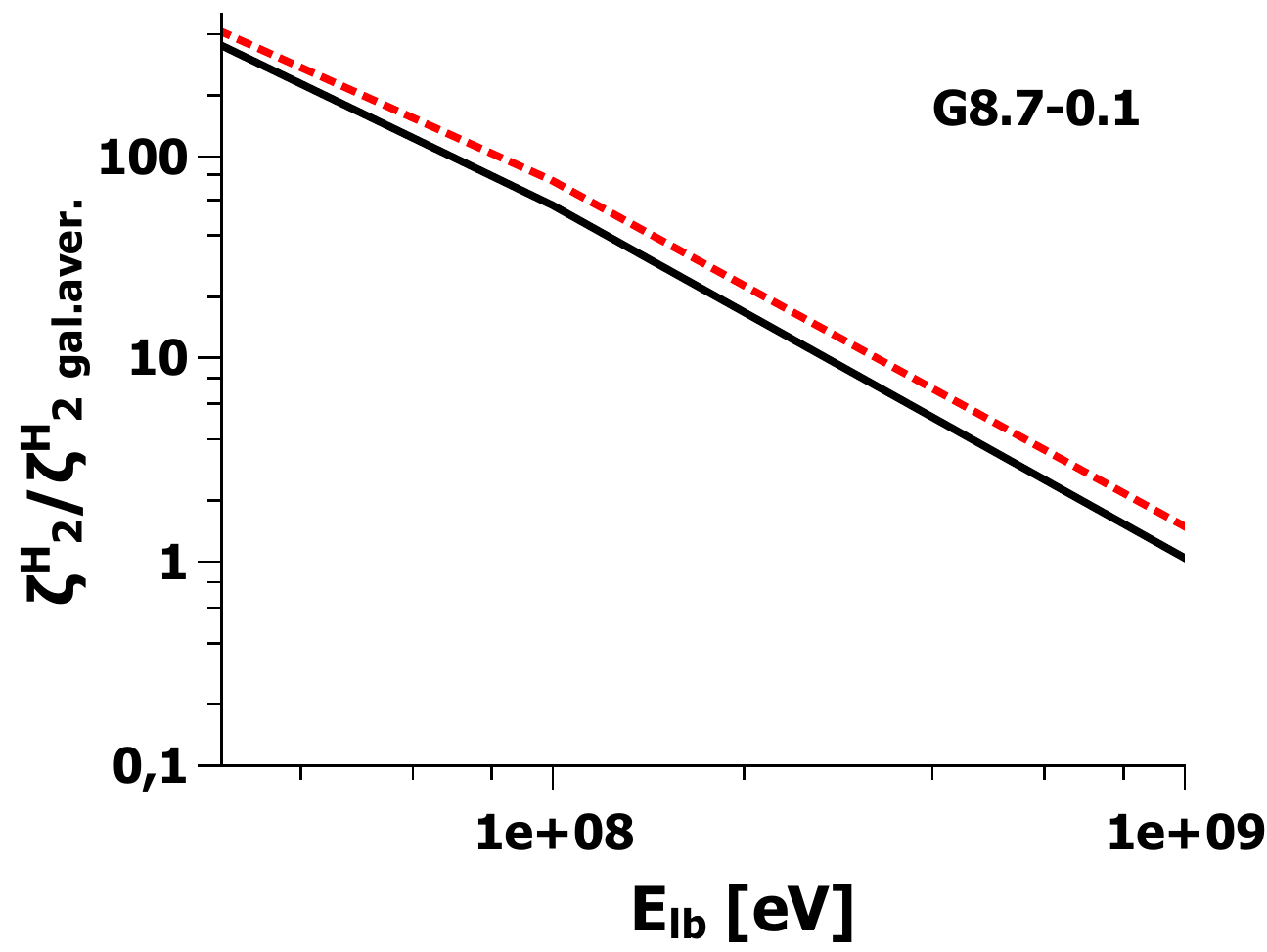} \\
		\end{tabular}
	\caption{Ionization rates versus lower break energy $E_{\rm lb}$ for different spectral indices below this break: $a~=~2.0$ (solid black line), $a~=~1.0$ (dotted red line).}
	\label{ion_plots}
\end{figure*}

Figure \ref{ion_plots} shows that even choosing a large value for the lower break energy would result in an ionization rate at least an order of magnitude greater than the Galactic average for molecular clouds, reported to be about $2 \times 10^{-16}\rm{~s}^{-1}$ by \cite{neufeld2010}, for at least two objects, namely W49B and 3C 391. The different lines refer to different values for the power law index $a$. As can be seen, the value of $a$ does influence the ionization rate, but only to a rather small extent. The ionization rate is much more sensitive to the choice of the lower break energy $E_{\rm lb}$. If $E_{\rm lb}~=~100$~MeV can be used, only the ionization rate for W51C would be lower than the Galactic average for molecular clouds, while the other objects would exceed this value for all spectral indices $a$. If even a value for $E_{\rm lb}$ as low as 30~MeV is suitable, the ionization rate for all objects would be greater than the Galactic average for molecular clouds. 

If even protons with a minimum energy of $E_{\min} = 2$~MeV could penetrate the cloud and thus contribute to the ionization, the ionization rates would increase further due to the maximum value of the ionization cross section at $E=100$~keV, as Fig.~1 in \cite{padovani2009} shows. According to \cite{indriolo2009}, this might be reasonable and would increase the ionization rate by 4 - 40$\%$ for $a$~$=$~$1$ and $E_{\rm lb}$~$=$~$30$~MeV. For larger values of $a$ and $E_{\rm lb}$, this enhancement of the ionization rate would be significantly lower.

The resulting ionization rates for each SNR known to be associated with a molecular cloud are given in Table \ref{results_ion}.

\begin{table*}
\centering{
\begin{tabular}{|c|c|c|c|}
	\hline
  object & $E_{\rm lb}$ & $\zeta^{\rm H_2}_{\rm gal}(a = 2.0)$ & $\zeta^{\rm H_2}_{\rm gal}(a = 1.0)$ \\
  \hline
  W51C & 1 GeV & 0.0259 & 0.0353 \\
  W51C & 100 MeV & 0.615 & 0.800 \\
  W51C & 30 MeV & 2.95 & 3.46 \\
  \hline
  W44 & 1 GeV & 0.752 & 1.04 \\
  W44 & 100 MeV & 21.0 & 27.4 \\
  W44 & 30 MeV & 116 & 139 \\
  \hline
  W28 & 1 GeV & 0.249 & 0.350 \\
  W28 & 100 MeV & 7.56 & 9.87 \\
  W28 & 30 MeV & 24.5 & 24.9 \\
  \hline
  IC443 & 1 GeV & 0.0323 & 0.0447 \\
  IC443 & 100 MeV & 1.54 & 2.01 \\
  IC443 & 30 MeV & 9.83 & 11.5 \\
  \hline
  W49B & 1 GeV & 1810 & 2580 \\
  W49B & 100 MeV & 66200 & 87200 \\
  W49B & 30 MeV & 385000 & 453000 \\
  \hline
  G349.7+0.2 & 1 GeV & 0.0688 & 0.0953 \\
  G349.7+0.2 & 100 MeV & 2.31 & 3.05 \\
  G349.7+0.2 & 30 MeV & 13.6 & 16.2 \\
  \hline
  CTB 37A & 1 GeV & 0.0385 & 0.0532 \\
  CTB 37A & 100 MeV & 1.23 & 1.62 \\
  CTB 37A & 30 MeV & 7.09 & 8.40 \\
  \hline
  3C 391 & 1 GeV & 9.28 & 16.4 \\
  3C 391 & 100 MeV & 91.4 & 121 \\
  3C 391 & 30 MeV & 567 & 663 \\
  \hline
  G8.7-0.1 & 1 GeV & 1.05 & 1.50 \\
  G8.7-0.1 & 100 MeV & 57.2 & 75.5 \\
  G8.7-0.1 & 30 MeV & 349 & 408 \\
  \hline
\end{tabular}
\caption{$\zeta^{\rm{H_2}} / \zeta^{\rm{H_2}}_{\rm{gal.~aver.}}$ for all objects, calculated for different spectral indices $a$ below the lower spectral break $E_{\rm lb}$ and different lower break energies $E_{\rm lb}$. \label{results_ion}}
}
\end{table*}

\section{Uncertainties\label{uncer}}

As mentioned above, since there is no observational data concerning the primary proton SED for energies below $E~\sim~$1~GeV, the spectral behavior in this low energy domain is unknown. However, assuming a positive spectral index $a~=~2$ is rather conservative and should offer a lower limit on the spectrum and therefore on the ionization rate. This is due to the fact that for an injection spectrum of $\propto~p^{-s}$, for a loss term of $\dot{p}~\propto~p^{-2}$ like ionization losses, would be modified as $\propto~p^{3-s}$, and the power law indices for the sources discussed are $s~=~1.5~$--$~2.45$~. However, this is likely to be the largest uncertainty. Furthermore, there is no precise estimate of the average density of the molecular clouds, so a value of $n_{\rm H}~=~100$~cm$^{-3}$ is assumed. This is also a major source of uncertainty. The primary proton SED scales linearly with the inverse value of $n_{\rm H}$ (see equation \ref{104}). Should the average density differ from the assumed value, then most likely it will be larger and therefore the primary proton SED and the ionization rate would decrease. Another critical parameter is the volume of the cloud. This is a particularly difficult aspect, since only the primary proton SED does depend on it (see equation \ref{103}), but not the resulting gamma spectrum (see equation \ref{102}). The radii used here refer to the whole SNRs and should therefore offer upper limits on the cloud volume, which would result in lower limits for the primary proton SEDs and thus the ionization rates. With this respect, our calculations therefore represent a conservative approach. The distance of the object from Earth enters the primary proton SED and the ionization rate as $a_{\rm p}~\propto~d_{\rm Earth-source}^{-1}$ (see equation \ref{104}), so the results are more sensitive on the lower break energy $E_{\rm lb}$ than on the distance.

Taking all the uncertainties discussed into consideration, still at least two sources, namely W49B and 3C~391, remain with an unusually large ionization rate at least one order of magnitude greater than the Galactic average for molecular clouds. Quite a few sources drop below Galactic average values in the most conservative scenario. This is not expected, since ionization at the accelerator itself is not likely to be lower than on average, but rather higher. Thus, we would rather expect the primary spectrum to extend toward lower energies or the interaction volume to be significantly smaller.

\section{Signatures\label{signatures}}

The enhanced ionization rate in the molecular clouds triggers a chemical network, forming a variety of ionized molecules in rotationally and vibrationally excited states, see e.g.\ \cite{black1998} and \cite{black2007}. Once these molecules are sufficiently abundant, their relaxation results in characteristic line emission.
Though in principle similar chemical signatures could be produced by X-ray ionization, X-rays have a much shorter penetration depth \citep{beskin2003accretion} and therefore are not capable of ionizing the cloud's interior as effectively as cosmic rays. This implies that the detection of chemical signatures similar to those shown in \cite{becker2011} would be a strong argument for a proton SED capable of producing these large ionization rates, which in turn would be a very strong argument for hadronic interactions as the dominant process in forming the detected gamma radiation in the GeV regime. Uncertainties do not allow the prediction of the exact value, but the statement that in general, an enhanced ionization rate is expected. Observations of the objects with e.g.\ Herschel and ALMA will give a better idea of exact values in the future. First results of such observations are summarized in section \ref{observations}.

In steady state the number densities of the transient ions O$^+$, OH$^+$, H$_2$O$^+$, and HeH$^+$ are expected to be proportional to $\zeta^{\rm{H}}$ \citep{herbst1973}. In fully molecular regions, where 
\begin{equation}
n(\rm{H}) \ll n(\rm{H_2})~,
\end{equation}
of a low ionization level 
\begin{equation}
x({\rm e}^-) < 10^{-5}~,
\end{equation}
competing processes almost do not interrupt the sequence of H-atom abstraction reactions. So nearly each ionization of hydrogen leads to the formation of H$^+_2$, H$^+_3$, OH$^+$, H$_2$O$^+$, and H$_3$O$^+$. The time-scales to achieve steady state are very short,
\begin{equation}
1 - 1000~\frac{\rm{cm}^{-3}}{n(\rm{H}_2)} \rm{~yr}~.
\end{equation}
These time-scales are shorter than the age of the SNRs dealt with here.

As stated also in \cite{becker2011}, the concrete ionization signatures to expect from the enhanced ionization rates cannot be predicted quantitatively yet. However, the statement that enhanced ionization signatures are to be expected can clearly be made. These signatures would be characteristic rotation-vibration emission lines from abundant ionized molecules as e.g.\ H$_2^+$ and H$_3^+$ (see \cite{becker2011} for an example of such a spectrum). The environments of SNR-MC systems are therefore considered to be optimal for a first time direct detection of H$_2^+$. Should such ionization signatures be detected in spatial correlation with the GeV gamma ray emission from an SNR associated with a molecular cloud, this would provide a strong hint at inelastic proton-proton scattering as the dominant process in forming these gamma rays.

\section{First experimental evidence\label{observations}}

In two sight lines in the IC443 complex, \cite{indriolo2010} found a large column density of H$_3^+$, $N({\rm H_3^+})=3\times10^{14}$~cm$^{-2}$, indicating an enhanced ionization rate of $\zeta^{\rm H_2}\sim2~\times~10^{-15}$~s$^{-1}$. This favors a less conservative calculation of the ionization rate in this region, as mentioned in section \ref{uncer}. 
Dense molecular gas associated with the W28 SNR shows evidence of
heating and chemistry driven by shock waves \citep{nicholas2011,nicholas2012}, but the
high excitation of ammonia molecules in one molecular core can also be interpreted
in terms of ion chemistry driven by an enhanced rate of ionization as outlined
in the next subsection.
A derivation of the ionization rate is given in subsection \ref{Radex W28}. In a molecular cloud of W51C, \cite{ceccarelli2011} derived an enhanced ionization rate by measurements of the DCO$^+$/HCO$^+$ ratio of $\zeta^{\rm H_2}\sim10^{-15}$~s$^{-1}$, which also favors a less conservative calculation of the ionization rate. 
These detections encourage additional observations toward the direction of the discussed SNR-MC systems, in particular toward the very promising candidates W49B and 3C 391.

\subsection{RADEX modeling for W28\label{Radex W28}}
The W28 SNR is a prominent example of an association of 
partly resolved HESS $\gamma$-ray sources and dense, shocked molecular
gas, as revealed by the molecular line observations of \cite{nicholas2011,nicholas2012}. In particular, the cm-wave inversion transitions of 
ammonia (NH$_3$) show peak intensities that coincide with the positions
of peaks in $\gamma$-ray emission and radio synchrotron emission. 
Conventionally NH$_3$ emission is thought
to probe dense molecular gas and the relative intensities of the inversion
transitions are considered to be good indicators of kinetic temperature. 
\cite{nicholas2011} identified a molecular cloud Core~2 that is associated with
the $\gamma$-ray source HESS J$1801-233$ and where the $(J,K) = (3,3)$ inversion
line of NH$_3$ is unusually intense compared with the $(1,1)$ and $(2,2)$ lines
of lower excitation. Moreover, they clearly detected emission in the $(6,6)$ 
line and weakly in the $(9,9)$ line at the same position. From a 
three-dimensional radiative transfer analysis of their most sensitive NH$_3$
spectra, \cite{nicholas2011} determined a best-fitting hydrogen number
density of $n_{\rm H}=10^{3.45}$ cm$^{-3}$ and a kinetic temperature $T=95$ K, 
but excluded the weak $(9,9)$ line from the analysis and noted that the model
underestimates the intensity of the $(6,6)$ line. This analysis makes the 
standard assumption that the excitation is controlled by inelastic collisions
of H$_2$ with NH$_3$ in competition with the radiative transitions. The NH$_3$
lines in Core~2 have unusually large line widths, indicating Doppler velocity
dispersions exceeding 5 km s$^{-1}$. Both the relatively high kinetic temperature and 
large velocity dispersion might result from the dynamical interaction of the
expanding SNR with a dense molecular cloud.
We have considered an alternative explanation of the high excitation of NH$_3$, 
which might apply in a region of enhanced cosmic-ray ionization rate. 
A chemical formation source for the highly excited inversion levels would naturally 
account for the superthermal line widths: the newly formed molecules would be 
translationally hot - they gain kinetic energy from the enthalpy change in the 
chemical formation process. 
If NH$_3$ is formed in a sequence of exoergic ion-molecule reactions that are
initiated by cosmic-ray ionization of hydrogen and/or nitrogen, then the 
formation process itself, mainly
$$ {\rm NH}_4^+ + e^- \to {\rm NH}_3 + {\rm H} \;\;\;, $$
leaves the product NH$_3$ molecules initially in highly excited rotational
states, which relax rapidly by submm-wave rotational transitions. However,
all steps in this radiative cascade that pass through rotational states $(J,K)$
with $J=K$, leave molecules stranded in the lower inversion level, because
all these levels are highly metastable. If the rate of formation of NH$_3$ is
high enough, compared with the rates of inelastic collisions, then the
formation process itself can account for observable populations of highly
excited states. Indeed, this mechanism of ``formation
pumping'' can mimic a collisionally excited component of molecular gas at 
a temperature $>100$ K, which would otherwise be needed to populate such 
metastable states as $(6,6)$ and $(9,9)$, which lie at energies $E(J,K)/k=
408$ K and 853 K above the ground state, respectively. 

We have calculated models of the non-LTE excitation of NH$_3$ in which the
formation and destruction processes are included together with the inelastic
collisions involving H$_2$ and $e^-$ and all relevant radiative processes, 
through use of the {\tt RADEX}
code as described by \cite{Radex2007}. 
There are too many free parameters and too many uncertainties in collisional
rates to permit a fully optimized model. However, the observed intensities of
the inversion lines of NH$_3$ are reproduced fairly well in a model at kinetic
temperature $T_k=80$ K with densities of H$_2$, $e^-$, and H$_3^+$ of 1000 cm$^{-3}$,
0.1 cm$^{-1}$, and 0.03 cm$^{-3}$, respectively. The fractional abundance of
ammonia relative to H$_2$ is $7.7\times 10^{-8}$ over a region of path length
$L=1.3\times 10^{19}$~cm = 4.2~pc, which corresponds to the extent of the
strong emission observed in the $(3,3)$ inversion line in cloud Core~2. The
H$_3^+$ ions are included in order to account for the relative amounts of
NH$_3$ in ortho-symmetry states, $(3,3)$, $(6,6)$, $(9,9)$, and in para-symmetry
states, $(1,1)$ and $(2,2)$, through reactive collisions that change the 
nuclear-spin symmetry of the molecule. In this model, the highly excited states
are populated largely by the formation process itself, and the inferred rate of
formation is $7.7\times 10^{-13}$ ammonia molecules cm$^{-3}$ s$^{-1}$. This is
balanced by a destruction rate of approximately $10^{-8}$ s$^{-1}$. These rates
would require that $\zeta^{{\rm H}_2}/\zeta^{{\rm H}_2}_{\rm gal} \sim 100$ and
thus imply a lower spectral break in the cosmic ray spectrum at $E_{\rm lb}\sim 
10$ MeV (see Table \ref{results_ion}). 

The high abundance and high excitation of NH$_3$ observed in molecular Core~2 
by \cite{nicholas2011} could be explained as the result of rapid formation
in an ion-driven chemistry. As discussed by Nicholas et al., the NH$_3$ might
result from shock-driven chemistry, which would also fit well with the other
tracers of molecular shocks that they describe. The clearest way to distinguish
between these two explanations would be observations of molecular ions like 
H$_3^+$ and H$_3$O$^+$, which would be expected to have greatly enhanced 
abundances if the cosmic ray ionization rate is unusually high.

\section{Conclusions and outlook\label{conclusions}}

In this paper we compute ionization rates for molecular clouds known to be associated with SNRs based on the assumption that the GeV gamma ray emission from these objects is due to neutral pion decay formed in proton-proton scattering in the molecular clouds. The computed ionization rates for at least two objects are above Galactic average for molecular clouds in the most conservative scenario. Therefore ionization signatures in the form of rotation-vibration line emission from molecular ions are likely to be detected from these two objects. The spatial correlation of the detection of GeV gamma rays on the one hand and rotation-vibration line emission from molecular ions on the other hand would strongly hint at the hadronic origin of the detected GeV gamma ray excess, explaining both detections by one population of cosmic ray particles.
However, there is the caveat that low energy CRs are efficient in ionizing, whereas high energy CRs are responsible for the gamma radiation. One can expect that low and high energy CRs are accelerated in the same objects, but an extrapolation of the high energy CR spectrum inferred from the gamma ray detections to low energies is not exempt from problems.
Still, recent observations hint at enhanced ionization rates in molecular clouds associated with SNRs, e.g.\ \cite{nicholas2011} and \cite{ceccarelli2011}, in support of the presented model.

In future work, differential propagation of the primary protons into the molecular cloud as well as the consideration of secondary ionization are planned to be implemented to offer more precise predictions concerning the ionization signatures to be expected.

\section{Acknowledgments\label{ack}}

 We would like to thank R.\ Schlickeiser and P.L.\ Biermann for helpful
 and inspiring discussions. We also want to thank the referee, Marco Padovani, for very helpful comments, which further improved the quality of this paper. JKB, SC and FS acknowledge funding from the DFG, Forschergruppe "Instabilities,
Turbulence and Transport in Cosmic Magnetic Fields" (FOR1048, Project BE
3714/5-1) and from the
Junges Kolleg (Nordrheinwestf\"alische Akademie der Wissenschaften und der
K\"unste). JHB is grateful to the Swedish National Space Board for
support. We further acknowledge the support by the
Research Department of Plasmas with Complex Interactions (Bochum).



\bibliography{lib_a_and_a}
\bibliographystyle{aa}
\end{document}